\def\kms{\hbox{km s$^{-1}$}}

\def\hii{H{\sc ii}\,}

\def\msun{M$_{\odot}$\,}
\def\jyb{Jy beam$^{-1}$}
\def\mjyb{mJy beam$^{-1}$}

\def\cm2{cm$^{-2}$}
\def\cm3{cm$^{-3}$}

\def\radec{\hbox{RA, Dec.(J2000)}}

\def\gra{$^{\circ}$}

\documentclass{aa} 
\usepackage{graphicx}
\usepackage{natbib}
\usepackage{graphicx}
\usepackage{txfonts}
\usepackage{pdflscape}
\begin{document} 

   \title{Triggered massive star formation associated with the bubble \hii region Sh2-39 (N5)  }

   \author{N. U. Duronea\inst{1}
        \and C. E. Cappa\inst{1,2}
	\and L. Bronfman\inst{3}
        \and J. Borissova\inst{4,5}
	\and M. Gromadzki\inst{4,5}
        }

\institute{Instituto Argentino de Radioastronom{\'{\i}}a, CONICET, CCT-La Plata, C.C.5., 1894, Villa Elisa, and CIC, Prov. de Bs. As. Argentina\ \email{duronea@iar.unlp.edu.ar} \and Facultad de Ciencias Astron\'omicas y Geof{\'{\i}}sicas, Universidad Nacional de La Plata, Paseo del Bosque s/n, 1900 La Plata,  Argentina \and  Departamento de Astronom{\'{\i}}a, Universidad de Chile, Casilla 36, Santiago de Chile, Chile \and Instituto de F\'isica y Astronom\'ia, Universidad de Valpara\'iso, Av. Gran Breta\~na 1111, Playa Ancha, Casilla 5030, Chile \and Millennium Institute of Astrophysics (MAS), Santiago, Chile}

   \date{Received September 15, 1996; accepted March 16, 1997}

 
  \abstract
   {}
   {Aiming at studying the physical properties of Galactic IR bubbles and to explore their impact in triggering  massive star formation, we perform a  multiwavelength analysis of the bubble \hii\ region Sh2-39 (N5) and its environs.  }
   {To analyze the molecular gas we use CO(3-2) and HCO$^+$(4-3) line data obtained with the on-the-fly technique from the ASTE telescope. To study the distribution and physical characteristics  of the dust, we make use of archival data from ATLASGAL, Herschel, and MSX, while  the ionized gas was studied making use of an NVSS  image. We use public WISE, Spitzer, and MSX point source catalogs to search for infrared candidate YSOs in the region. To investigate the stellar cluster [BDS2003]6 we use IR spectroscopic data obtained with the ARCoIRIS  spectrograph, mounted on Blanco 4-m Telescope at CTIO, and new available IR Ks band observations from the VVVeXtended ESO Public Survey (VVVX). }
   {The new ASTE observations allowed the molecular gas component in the velocity range from 30 \kms\ to 46 \kms, associated with Sh2-39,  to be  studied in detail. The morphology of the molecular gas  suggests that the ionized gas is expanding against its parental cloud. We have identified four  molecular clumps, that were likely formed by the expansion of the ionization front, and determined some of their physical and dynamical properties. Clumps having HCO$^+$ and 870 $\mu$m counterparts show evidence of gravitational collapse.   We identified several candidate  YSOs across the molecular component. Their spatial distribution, as well as the fragmentation time derived for the collected layers of the molecular gas, suggest that massive star formation might have been triggered by the expansion of the nebula via the collect and collapse mechanism. The spectroscopical distance obtained for the stellar cluster [BDS2003]6, placed over one of the collapsing clumps in the border of the \hii\ region,  reveals that this cluster is physically associated with the neabula and gives more support to the triggered massive star formation scenario.     A  radio continuum data analysis indicates that the nebula is older and expands at lower velocity than typical IR Galactic bubbles.  }
   {}


   \maketitle
%

\section{Introduction} 

 Although massive stars formation processes  are  still under debate \citep{zin07,mckee07}, it is  believed that they are born inside dense  ($n$ $\sim$ 10$^{3-8}$ cm$^{-3}$), small (R $\sim$ 0.1 - 5 pc), and massive (M $\sim$ 100 - 1000 \msun) clumps of molecular gas and dust with high optical  extinction. Understanding the formation and early stages of massive stars requires then a deep knowledge of the original physical conditions of the regions  where they are born, as well as their surroundings.

 It is well known that a considerable number of massive stars in our Galaxy can be formed  by the action of \hii regions over their parental molecular environment through processes like the {\it collect-and-collapse} mechanism (C$\&$C; \citealt{elm77}) or  the {\it radiative driven implosion} process (RDI;  \citealt{lela94}).   In recent years, the triggered star formation process, especially the C$\&$C mechanism, has been studied extensively at the edges of many bubble-shaped  \hii\ regions such as Sh2-104, RCW 79, Sh2-212, RCW 120, Sh2-217, Sh2-90, Gum 31, and S 24 \citep{deh03,deh08,deh09,zav06,zav10,bra11,sam14,du15,ca16}. Dense molecular clumps placed along  the border of  Galactic bubble  \hii regions are therefore among the most likely sites for stellar births, and hence, where to look for early stages of massive star formation.    Keeping this in mind, it is instructive to study the interstellar medium  adjacent to these objects  since they  can provide important information on the molecular environment where massive stars can be formed. 

 As  part of a project aimed at studying  the  physical properties of IR bubbles and to better understand their influence  in triggering massive star formation,    we have selected the northern nebula \hbox{Sh2-39} \citep{sha59},  a  bubble nebula  that was partially imaged in the {\it Spitzer} GLIMPSE survey,  listed as N5 by \citet{chu06},  and  barely  studied by \citet{bea10}. The selection of this object  was made taking into account the detection of the  \hbox{CS(2-1)} line  from the survey by \citet{bro96}, ensuring the existence of  high density molecular gas, a necessary condition for the massive star formation.     

\begin{figure*}
   \centering
   \includegraphics[width=470pt]{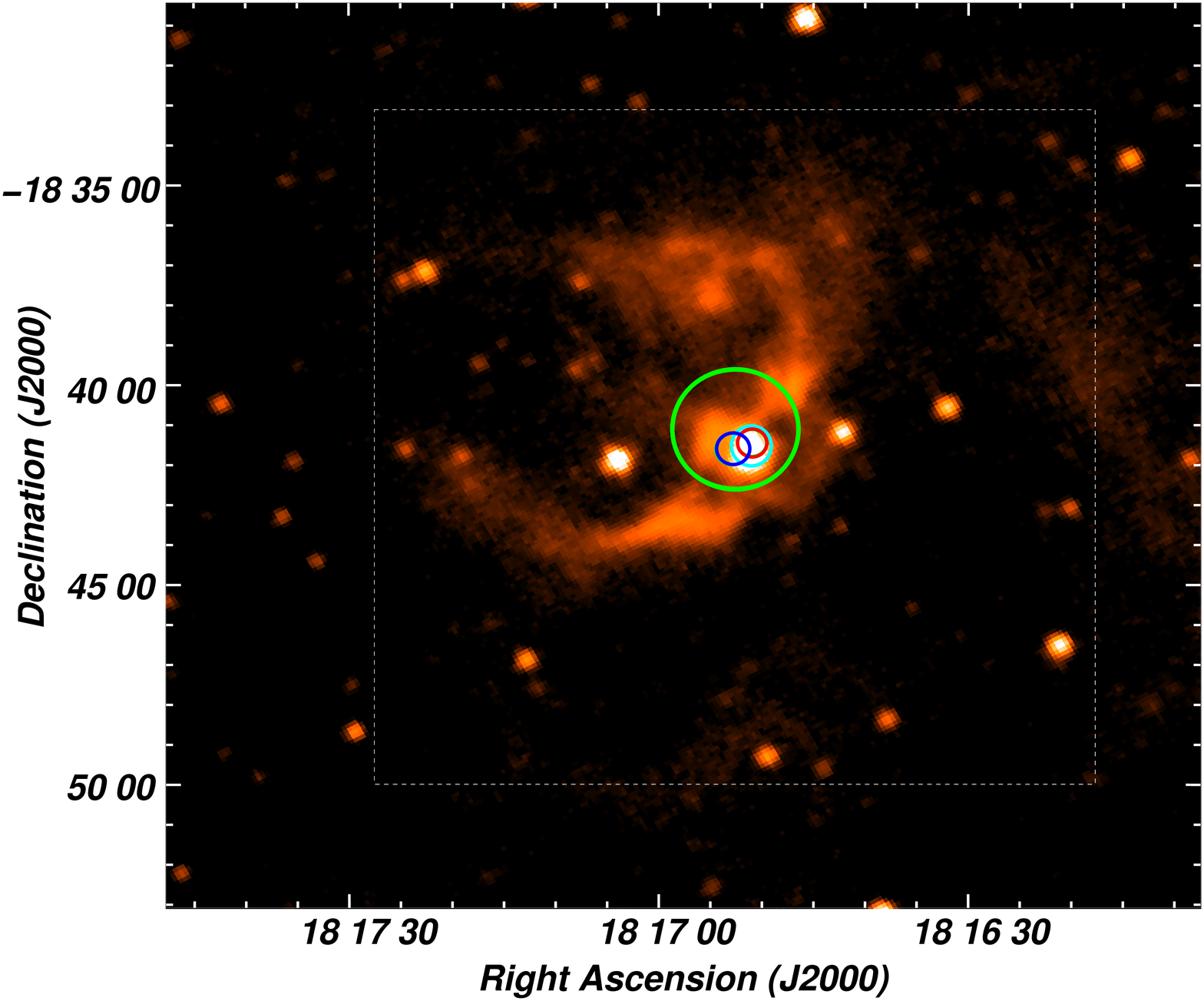}
   \caption{MSX-A  (8.28 $\mu$m) emission of the IR bubble nebula N5 and its environs. The white dashed square indicates the area observed with the ASTE telescope. The position and beam size of the CS(2-1) line observation from \citet{bro96} is shown with the blue circle, while red and green circles show the position and beam size of the  observations from \citet{ur11} and \citet{loc89}, respectively  (see text).      The approximate location and size of the cluster [BDS2003]6 is also shown with the cyan circle.  }
   \label{msx}%
    \end{figure*}
 
Here we present the very first multiwavelength study of \hbox{Sh2-39}  and its environs. We have carried out observations of the CO(3-2) and HCO$^+$(4-3) lines with the ASTE telescope, aiming at studying in detail the spatial distribution and physical characteristics of the molecular gas associated with the nebula. To analyze  the  properties of  the  dust  in the nebula  and its surroundings we use images from the  ATLASGAL and {\em Herschel} databases. To account for the interaction between the ionized and molecular gas, and to determine the  evolutionary status of the \hii region we use radio continuum data from the NVSS archives.   This analysis  will also allow  to  identify spots of  dense molecular gas and dust where the massive star formation is developing, giving special attention to the central region corresponding to the source  IRAS 18139-1842, where star formation processes seem to be very active.  In addition, we look for signatures of star formation in the surroundings using archival photometric data and we investigate a probable scenario of triggered star formation at the border of the nebula.  We also study the characteristics  of an identified  stellar cluster projected onto the border of the bubble using  new IR data. This will provide some additional  information that will help to better understand the star formation activity in the nebula.

N5 is an open  IR bubble of  about 8\arcmin\ in diameter centered at \radec\ =\ 18$^h$17$^m$02$^s$, $-$18\gra40\arcmin19\arcsec, coincident with the \hii region Sh2-39,  and located at a distance of 4.1 kpc \citep{bro96,fau04}.  In Fig.~\ref{msx} we present a  8.28 $\mu$m (MSX-A) emission image of the nebula, which has a  typical arc-shaped morphology in the mid-IR wavelengths. At  wavelengths of about $\sim$ 8 $\mu$m  most of the emission originates in strong features of PAH molecules,  which are considered to be good tracers of warm UV-irradiated photodissociation regions (PDR; \citealt{ht97}).  Since these complex molecules are destroyed inside the ionized gas of an \hii\ region, they indicate the limits of the ionization front and delineate the boundaries of the bubble nebula, giving a  glimmer of the location of the parental molecular gas against which N5 is likely expanding. 
 The  open morphology of the IR nebula suggests that Sh2-39 is the classical  \hii\ region density bounded toward the outer side (eastwards) and  ionization bounded toward the molecular cloud (westwards).  

     The nebula is  believed to be linked to the IR sources IRAS 18139-1842 and IRAS 18139-1839, both with characteristics of massive young stellar objects (see Sect. 6). The \hbox{CS(2-1)} line was detected by  \citet{bro96} at a velocity\footnote{All the velocities in this work are referred to the LSR} of 40.2 \kms\  in the direction of IRAS 18139-1842. NH$_3$ line and H$_2$O maser emission (believed to be excellent indicators of early stages of massive star formation) were also detected towards this IRAS source \citep{su07,ur11}, while  radio recombination line observations (H87$\alpha$ and H88$\alpha$)  show emission at the same position at a velocity of 41 \kms\ \citep{loc89} (see Fig.~\ref{msx}),  which indicates the existence of high density ionized gas at the same radial velocity than the molecular gas.  The radio source G012.4317-01.1112, presumably associated with IRAS 18139-1842 was studied by \citet{urq09} who classified the source as a classical HII region.      

Regarding the stellar content in the region of the nebula, using the 2MASS catalog \citet{bica03}   identified the IR cluster candidate [BDS2003]6, of about 0\arcmin.8 in size, coincident with the position of IRAS 18139-1842 (see Fig. \ref{msx}). The presence of this stellar cluster projected onto the IR emission  is another strong evidence for  massive star formation in the border of the bubble.

\section{Observations and complementary data}

\subsection{Molecular data}

The molecular observations were carried out in August 2015 with the 10 m Atacama Submillimeter Telescope Experiment (ASTE;  \citealt{ez04,ez08}). We used DASH345, a two-sideband single-polarization heterodyne receiver, tunable in LO frequency range from 327 GHz to 370 GHz at  observable frequency range from 321 GHz to 376 GHz. The XF digital spectrometer was set to a bandwidth and spectral resolution of 128 MHz and 125 KHz, respectively. The spectral velocity resolution was 0.11 \kms, the half power beamwidth (HPBW) is $\sim$ 22'', and the main beam efficiency ($\eta_{\rm mb}$) is 0.65. Observations were made using the on-the-fly (OTF) mode with two orthogonal scan directions along RA and Dec.(J2000) centered on \radec = (18$^h$17$^m$02.1$^s$, $-$18\gra40\arcmin19\arcsec). We observed simultaneously the lines CO(3-2) (345.796 GHz) and HCO$^+$(4-3) (356.734) in  a region of $\sim$ 17$'$ $\times$ 17$'$ (see Fig.~\ref{msx}). The spectra were reduced with  NOSTAR\footnote{http://alma.mtk.nao.ac.jp/aste/guide/otf/reduct-e.html} using the standard procedure.

\subsection{IR spectroscopic data}

The brightest star projected at the center of [BDS2003]\,6  (2MASSJ18165113-1841488) was observed on August 2016  with Astronomy Research using the Cornell Infra Red Imaging Spectrograph (ARCoIRIS), a cross-dispersed, single-object, longslit, infrared imaging spectrograph, mounted on Blanco 4-m Telescope, CTIO. The spectra cover a simultaneous wavelength range of 0.80 to 2.47 $\mu$m, at a spectral resolution of about 3500 $\lambda$/$\Delta \lambda$, encompassing the entire zYJHK photometric range. The spectrum was taken with 480 sec integration time, at 1.03 average airmass. The HD163336 tellluric A0 V standard is observed immediately after target.  The basic steps of the reduction procedure are described in \citet{chene12,chene13}. We used the corresponding pipeline\footnote{http://www.ctio.noao.edu/noao/content/Data-Reduction}.

\subsection{Archival data}

The data described above  were complemented with several archival data sets:

\begin{itemize}

\item Infrared data: {\it a)} images of ATLASGAL at 870 $\mu$m (345 GHz)  \citep{sch09}. This survey covers the inner Galactic plane, $l$ = 300\gra\ to 60\gra, $|b|$ $\leq$ 1.\gra5, with a rms noise in the range 0.05 - 0.07 \jyb.  The {\it Large APEX BOlometer CAmera} (LABOCA) used for these observations, is a 295-pixel bolometer array developed by the Max-Planck-Institut fur Radioastronomie \citep{sir07}. The beam size at 870 $\mu$m is 19\farcs2.  {\it b)} The new project ``The VVV eXtended ESO Public Survey'' (VVVX; Hempel et al. 2017, in prep.) was approved as an extension of the ``VISTA Variables in the Via Lactea'' (VVV; \citealt{minn10,saito12}) to enhance its long lasting legacy. The VVVX Survey will extend the VVV time-baseline and will cover  additional to VVV 1700 sq/degrees in the Southern sky (7 h $<$ RA $<$ 19 h).  The VVVX started in 2016 and the region of N5 was observed in Ks band in August 2016. We have retrieved 10 images from Cambridge Astronomical Survey Unit (CASU) VIRCAM pipeline \citep{irw04}, each of them with 4 sec exposure time covering area of 11$'$ $\times$ 11$'$. The images were combined using standard IRAF procedures. {\it c)} Images from the {\em Herschel}\footnote{{\em Herschel} is an ESA space observatory with science instruments  provided by European-led Principal Investigator consortia and with important participation from NASA (http://www.cosmos.esa.int/web/herschel/science-archive)} Infrared GALactic (Hi-GAL) plane survey key program \citep{mol10}. We used images from the SPIRE archive data \citep{grif10}   at  250\,$\mu$m and 350\,$\mu$m, with beam sizes of 18$''$ and 25$''$, respectively. We retrieved images processed at level 2.5 from the NASA/IPAC database\footnote{http://irsa.ipac.caltech.edu/frontpage/}. 
{\it d)} Infrared data at 8.28 $\mu$m retrieved from the  Midcourse Space Experiment (MSX)\footnote{http://irsa.ipac.caltech.edu/Missions/msx.html} \citep{p01}. The image has a spatial resolution of 20$''$. \\

\item Radio continuum image at 1.4 GHz  obtained from the NRAO Very Large Aray (VLA) Sky Survey\footnote{http://www.cv.nrao.edu/nvss/} (NVSS; \citealt{con98}). The images have a spatial resolution of 45$''$ and a rms noise of $~$0.45 \mjyb\  (Stokes I).  \\


\item To investigate the existence of candidate young stellar objects (YSOs) projected onto the region we analyzed infrared point sources from the MSX survey \citep{p01}, the Galactic Plane Survey (GPS) of the the UKIRT Infrared Deep Sky Survey (UKIDSS) \citep{lucas08}, the Spitzer survey (\citealt{fazio04}), and the  Wide-field Infrared Survey Explorer (WISE, \citealt{wright10}).

\end{itemize}


\section{Molecular gas analysis}

\subsection{Spatial distribution of the molecular gas and identification of clumps}

\begin{figure}
   \centering
   \includegraphics[width=260pt]{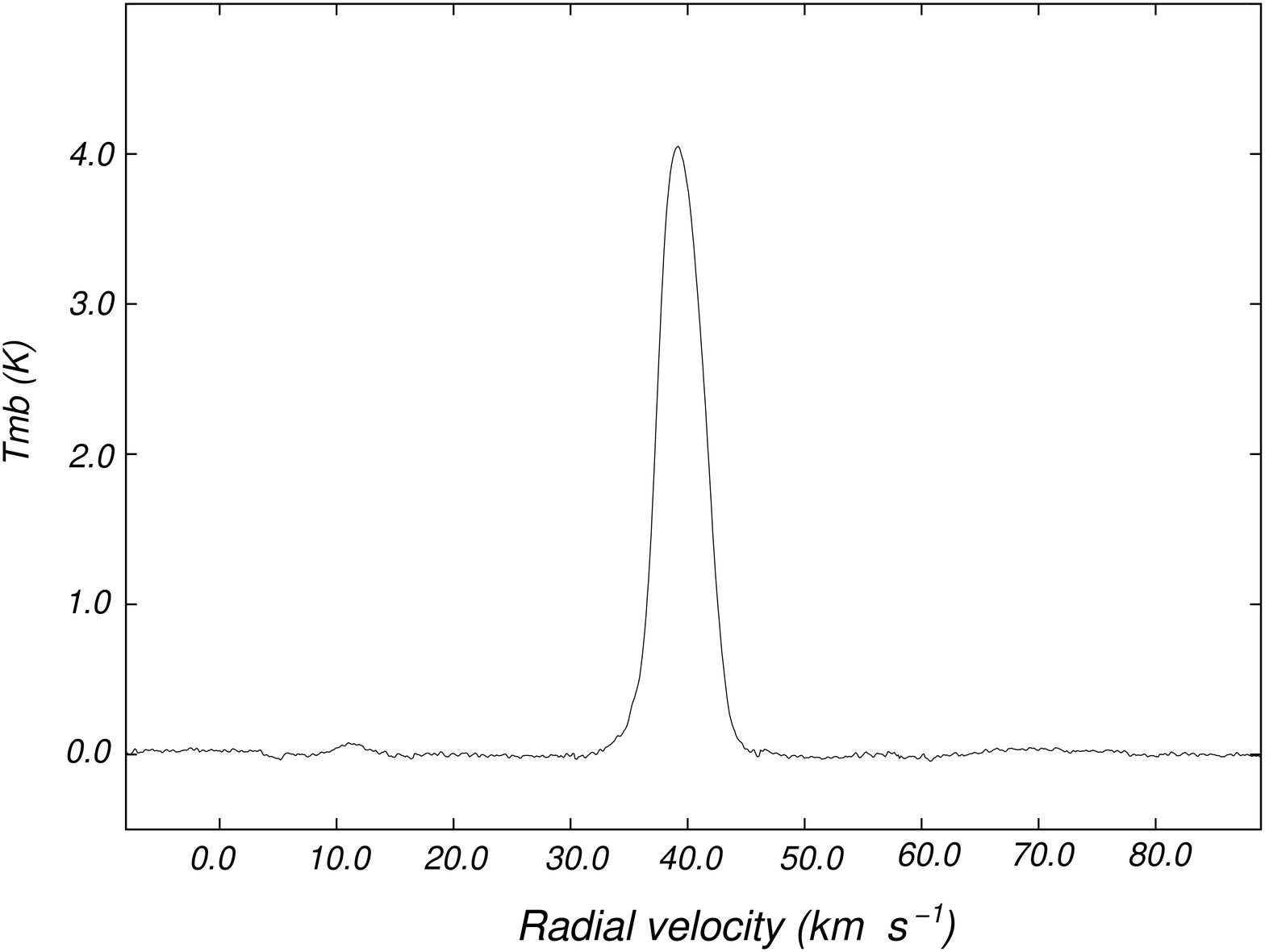}
   \caption{Averaged spectrum obtained within a region of $\sim$ 12$'$$\times$12$'$ centered on \radec\ =\ 18$^h$16$^m$53$^s$, $-$18\gra40\arcmin50\arcsec. }
   \label{promedio}%
    \end{figure}

In Fig.~\ref{promedio} we show the CO(3-2) spectrum averaged over a region $\sim$ 12$'$$\times$12$'$ in size  around the center of the nebula  (\radec\ =\ 18$^h$17$^m$02$^s$, $-$18\gra40\arcmin19\arcsec ). The molecular emission appears concentrated in only one component between $\sim$ 30 \kms\ and 46 \kms\ with a peak emission around 39.5 \kms. This velocity is almost coincident with the peak velocity of the CS(2-1) line reported  by \citet{bro96} and the HH87$\alpha$ recombination line reported by \citet{loc89} towards IRAS 18139-1842.
    
\begin{figure*}
   \centering
   \includegraphics[width=520pt]{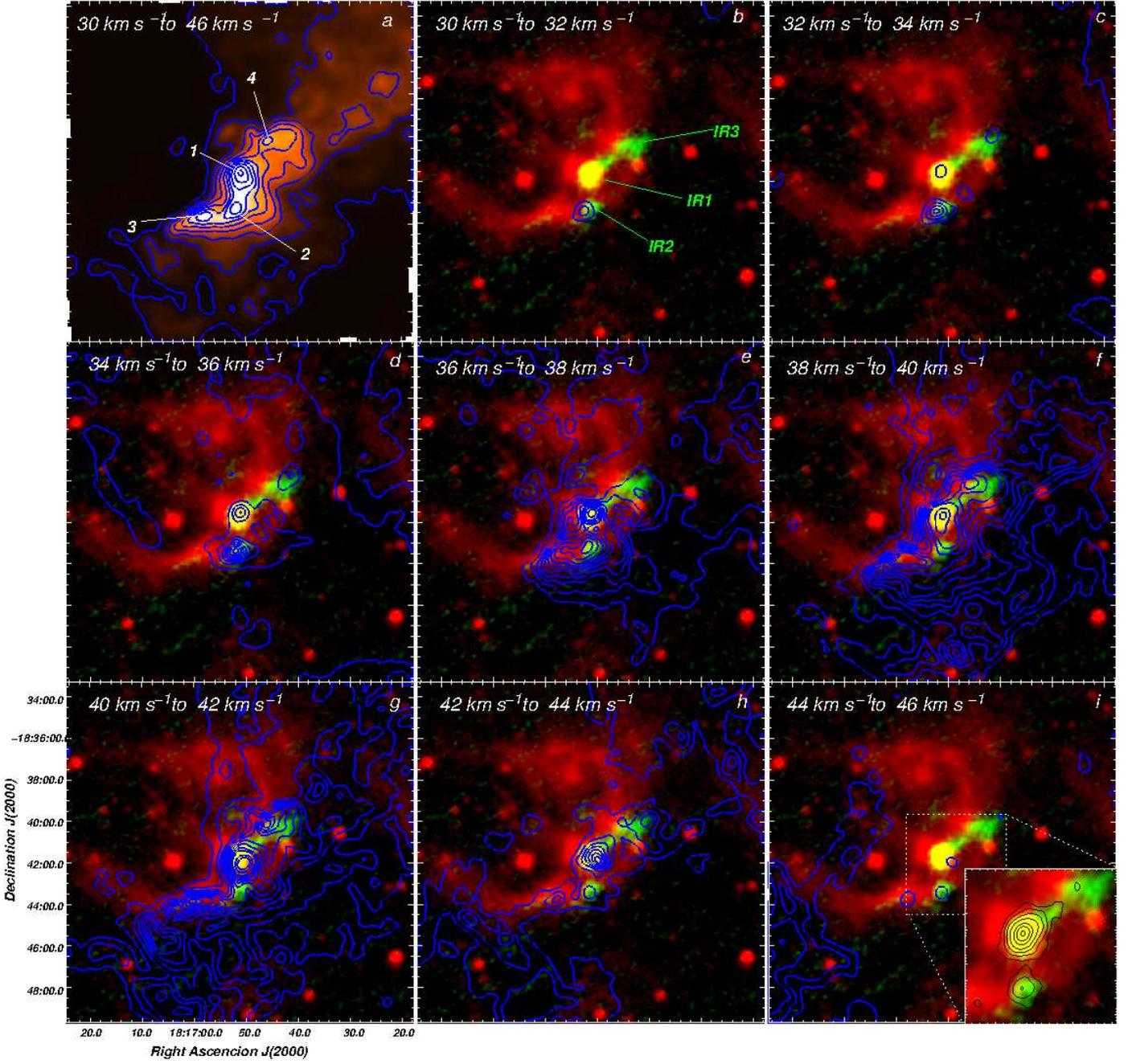} 
   \caption{Panel {\it a}: CO(3-2) emission in the velocity range from 30 to 46 \kms. Contour levels go from 0.3 K \kms\ ($\sim$ 20 rms) to 3.3 K \kms\ in steps of 0.5 K \kms\ and from 4 K \kms\ in steps of 1 K \kms.  White numbers indicate the identified molecular clumps (see text). Panels { \it b} to {\it i}: Channel maps of CO(3-2) in intervals of 2 \kms (blue contours) superimposed on the 8.28 $\mu$m MSX-A emission (red colorscale) and 870 $\mu$m ATLASGAL emission (green colorscale). Contours go from 0.2 K \kms\ ($\sim$ 8 rms) to 9.2 K \kms\ in steps of 1 K \kms, and from 11 K \kms\  in steps of 2 K \kms. Infrared sources identified in the 870 $\mu$m emission  (see Sect. 4.2) are indicated in panel {\it b} in green.  A zoom of the central region with the emission of the HCO$^+$(4-3) line (black contours) in the velocity interval from 36.5 \kms\ to 42.4 \kms\ is shown in the lower right corner of panel {\it i}. HCO$^+$ contour levels are 0.07 ($\sim$ 8 rms), 0.12, 0.21, 0.35, 0.6, 0.8 and 1.0 K \kms.   }
   \label{mosaico}%
    \end{figure*}

  In the panel {\it a} of Fig.~\ref{mosaico} we show the integrated CO(3-2) emission distribution in the mentioned velocity range. As expected, the molecular gas appears concentrated towards the western and southern borders  of the IR nebula, while no significant molecular emission is detected towards the eastern and north-eastern borders. This is  in line with the proposed scenario of an \hii\ region bounded by density to the east and by ionization to the west.  The morphology and location of the molecular gas and radio continuum emission (see Sect. 5)  suggest that the ionized gas is expanding against the molecular cloud, probably in a  blister-type \hii\ region scenario \citep{i78}. The morphology, distribution and location of the molecular gas with respect to the near IR emission is similar to many other in Galactic IR bubbles found in the literature.

   Although morphological  evidence suggests that the molecular cloud is expanding,  we performed several tests to detect expanding motions in the molecular gas (not shown here) with no conclusive results. This very likely indicates that the bubble is evolved, or else is expanding at a low rate. This scenario will be further tested in Sect. 7.2 by analyzing the properties of the ionized gas.      

The molecular emission is not uniformly distributed, and several condensations can be distinguished in the whole structure. These condensations,  referred  hereafter to  as 'clumps'   \citep{blitz93}, must be identified and their physical properties must be estimated if a study of the molecular gas in Sh2-39 is to be carried out. To identify the CO clumps and to  unambiguously ascertain their geometrical and physical properties  we made use of the {\it clumpfind} algorithm \citep{will94}. We were left only with  the  brightest clumps which are, in turn,  those adjacent to the bubble. These clumps  are the places where star formation seems to be  taking place (see Sect. 6) and were very likely formed in the collected layers of the molecular gas as a result of the expansion of the nebula against the parental molecular cloud.  The location of the clumps, labeled from 1 to 4, is depicted in Fig.~\ref{mosaico}a and their observational  parameters are indicated in Table \ref{tabla-medidas}. The area of the clumps in the CO emission ($A^{\rm CO}_{\rm clump}$;  Col. 4 in Table \ref{tabla-medidas}) was estimated with the {\it clumpfind} algorithm. The peak temperatures ($T^{\rm CO}_{\rm peak}$; Col.5 in Table \ref{tabla-medidas}) and the velocity of the CO emission peak (v$^{\rm CO}_{\rm peak}$;  Col. 6 in Table \ref{tabla-medidas}) were estimated by gaussian fitting in spectra obtained in the direction of the maximum emission of each clump. The line width of the CO line for each clump ($\Delta{\rm V}^{\rm CO}$; Col.7 in Table \ref{tabla-medidas}) was obtained as 2 $\times$ $\sqrt{2\ {\rm ln}2}$ $\times$ $\sigma_{\rm gauss}$.  


\begin{table*}
\caption{Observed parameters of the CO and HCO$^+$  clumps}
\centering
\begin{tabular}{ccccccccccc}
\hline
clump  &  RA & Dec. & $A^{\rm CO}_{\rm clump}$ & $T^{\rm CO}_{\rm peak}$ & v$^{\rm CO}_{\rm peak}$ & $\Delta$V$^{\rm CO}$ & $A^{\rm HCO+}_{\rm clump}$ & $T^{\rm HCO^+}_{\rm peak}$ & v$^{\rm HCO+}_{\rm peak}$ & $\Delta$V$^{\rm HCO+}$ \\
       & ($^h$\ $^m$\ $^s$) & (\gra\ \arcmin\ \arcsec) &  (10$^{-7}$ ster)   &   (K)                &     (\kms)       &      (\kms)                 &  (10$^{-7}$ ster)           &     (K) &   (\kms) &     (\kms)       \\ 
\hline 
\hline
&&&&&&&\\  
1  &    18  16  50.8    &    $-$18  41  40       &  3.9       &  10.6  &  38.3   & 6.5  & 1.9   &  1.1   & 40.1   & 3.3  \\
2  &   18  16  51.8     &   $-$18  43  18       &  6.7       &  6.3   &  38.7   & 6.6  & 0.8   &  0.3   & 38.5   & 2.9  \\
3  &   18  16  57.9     &   $-$18  43  39       &  4.5       &  7.2   &  35.1   & 5.9  &  -     & -       & -   &   \\
4  &   18  16  45.8     &   $-$18  40  11       &  6.1       &  8.5   &  35.5   & 3.8  &  -     & -       & -   &   \\
\hline
\end{tabular} 
\label{tabla-medidas}
\end{table*}

\begin{table*}
\caption{Physical properties derived for  the CO and HCO$^+$ clumps}
\label{tabla-propiedades}
\centering
\begin{tabular}{ccccccccc}
\hline
clump  & $R^{\rm CO}_{\rm D}$  & $N({\rm H_2})$ & $M_{\rm int}(\rm H_2)$   &  $n_{\rm (H_2)}$       &  $R^{\rm HCO^+}_{\rm D}$ & $N({\rm HCO^+})$ &  $M_{\rm LTE}$  &  $M_{\rm vir}$    \\
       &  (pc)             &  (10$^{22}$ cm$^{-2}$)   &   (10$^3$ M$_{\odot}$)  &   (10$^3$ cm$^{-3}$)  &  (pc)               &   (10$^{12}$ cm$^{-2}$)    & (10$^{3}$ \msun) & (10$^3$ \msun) \\  
\hline
\hline
1  &  1.4     &  1.3  & 2.0    & 3.3      &  1.0      &  1.0 - 2.3    & 1.3 - 3.2   & 1.3 -2.1 \\
2  &  1.9     &  0.9  & 2.2    & 1.5      &  0.6      &  0.4 - 1.0    &  0.5 - 1.5 &  0.6 -0.9 \\
3  &  1.5     &  0.8  & 1.4    & 1.8      &   -     & -    &  -    &    -      \\
4  &  1.8     &  0.9  & 1.9    & 1.6      &   -     & -    &  -    &     -  \\
\hline
\end{tabular} 
\end{table*}

 In order to provide a better visual display of the dynamics of the molecular gas component, in panels {\it b} to {\it i} of Fig.~\ref{mosaico}  we show the channel maps of CO(3-2) in velocity intervals of 2 \kms , overlaid onto the  8.28 $\mu$m (in red)  and 870 $\mu$m (in green)  emissions. In the velocity interval from 30 \kms\ to 34 \kms\ (panels {\it b} and {\it c}) clump 2 becomes noticeable, achieving its maximum emission temperature in the velocity interval from 36 \kms\ to 38 \kms\ (panel {\it e}). This clump is detected up to  a velocity of 39 \kms\ (not shown here) where it is mixed with  the emission of clumps 3 and 1.  Clumps 1 and 4 become detected at velocities between 32 \kms\ to 36 \kms\  (panels {\it c} and {\it d}). In this velocity interval a faint extended molecular emission appears bordering the western and north-western side of the nebula. This structure will be deteced up to  a velocity of $\sim$ 42 \kms. Clump 1 and 4 achieve their  maximum emission temperature in the velocity interval from 40 \kms\ from 42 \kms\ (panel {\it g}) and from 38 \kms\ to 40 \kms\  (panel {\it f}), respectively. Both clumps are detected till a velocity of $\sim$ 45 \kms. Clump 3 is noticed in the velocity interval from 36 \kms\ to 44 \kms\ (panels {\it e} to {\it h}), achieving its maximum emission temperature in the velocity interval from 38 \kms\ to 40 \kms\ (panel {\it f}). In the velocity interval from 38 \kms\ to 46 \kms\ (panels {\it f} to {\it i}) the south-eastern border of the nebula is surrounded by a filamentary molecular structure detected up to  a velocity of $\sim$ 47 \kms. Unlike IRAS 18139-1842, which seems to be pysically related with clump 1,  IRAS 18139-1839 does not seem to have a molecular gas component associated.  No substantial molecular emission is detected either towards the center or the east of the nebula suggesting a density gradient in the region. 

In the bottom right corner of Fig.~\ref{mosaico}{\it i} we display an enlargement of the central region showing the emission of the \hbox{HCO$^+$(4-3)} line (black contours) in the velocity range from 35.5 \kms\ to 42.4 \kms. Two molecular structures (or clumps) are detected with centers at  \radec\ =\ 18$^h$16$^m$51.4$^s$, $-$18\gra41\arcmin36\arcsec and  \radec\ =\ 18$^h$16$^m$51.6$^s$, $-$18\gra43\arcmin09\arcsec. Their location and size suggest that these clumps are the HCO$^+$ counterparts of CO clumps 1 and 2. The detection of the HCO$^+$(4-3) line, as well as 870 $\mu$m emission,  indicate that these clumps have high densities making them as potential good candidates sites to form massive stars. The location of the stellar cluster [BDS2003]6 in the direction of clump 1 (see Fig.~\ref{msx}) gives more support to this scenario.  

For the case of the HCO$^+$ clumps, their areas ($A^{\rm HCO+}_{\rm clump}$; Col. 8 in  Table \ref{tabla-medidas}) were estimated using the contour level of 0.07 K \kms ($\sim$ 8 rms).  The peak temperature of the HCO$^+$ clumps  ($T^{\rm HCO^+}_{\rm peak}$; Col. 9, in  \ref{tabla-medidas}) and the velocity of the HCO$^+$ emission peak (v$^{\rm HCO+}_{\rm peak}$;  Col. 10 in Table \ref{tabla-medidas})    were estimated by gaussian fitting in spectra obtained towards  the maximum emission of each clump. The line width of the HCO$^+$ line ($\Delta{\rm V}^{\rm HCO^+}$) is indicated in Col. 11 in Table~\ref{tabla-medidas}.


\subsection{Physical properties of the clumps}

  The deconvolved effective radius of the clumps derived from the CO and HCO$^+$ lines, $R^{\rm CO}_{\rm D}$ and $R^{\rm HCO+}_{\rm D}$  (Cols. 2 and 6  in Table  \ref{tabla-propiedades}) were calculated as
\begin{equation}
  R_{\rm D} =   \sqrt{R^2_{\rm eff} - {\rm HPBW}^2/4},
\end{equation} 
where  $R_{\rm eff}$  is the effective radii of the clumps ($R_{\rm eff}$ = $\sqrt{A_{\rm clump}/\pi}$) calculated in both lines and HPBW is the half-power beam width of the instrument.

To estimate the mass of the clumps we use the relation between the H$_2$ column density ($N({\rm H_2})$; Col. 3 in Table~\ref{tabla-propiedades}) and the CO integrated emission
\begin{equation}
  N({\rm H_2})\ =\ X\ \times \int{T_{\rm  mb}({\rm CO})}\ d{\rm v}, 
\end{equation} 
where $X$ is an empirical factor that has been shown to be roughly constant for Galactic molecular clouds. For  the \hbox{$^{12}$CO(1-0)} line the $X$ value is about 1.9 $\times$ 10$^{20}$ cm$^{-2}$(K \kms)$^{-1}$ \citep{strong96}. Since we use the integrated intensity emission of the CO(3-2) line we need to adjust the value of $N({\rm H_2})$  using a correcting factor of $\sim$  0.7 \citep{oka12}. Then, the total integrated hydrogen mass ($M_{\rm int}({\rm H_2})$, Col 4 in Table \ref{tabla-propiedades})  is calculated using
\begin{equation}
  M_{\rm int}(\rm H_2)\ =\  \ \mu\ \ {\it m}_H\ \ \sum\ \ A_{\rm clump}     \ \frac{{\it N}(\rm H_2)}{0.7}\ \ {\it d}^2 ,
\label{eq:masa}
\end{equation} 
where   $\mu$ is the mean molecular weight, which is  assumed to be equal to 2.8 after allowance of a relative helium abundance of 25\% by mass,  $m_{\rm H}$ is the hydrogen atom mass,  $A_{\rm clump}$    is the area subtended by the CO clump (see Table~\ref{tabla-medidas}), and  $d$ is the kinematical distance (assumed to be 4.1 kpc). The volume density ($n_{\rm (H_2)}$, Col. 5 in Table \ref{tabla-propiedades}) was derived assuming spherical geometry for each clump.  

\begin{table*}
\caption{Properties derived for the submillimeter sources}
\centering
\begin{tabular}{cccccccccccc}
\hline
source & RA & Dec.  & $S_{870}$   & $I_{870}$    & $R_{\rm IR}$ & $\bar{T}_{\rm dust}$ &  $T^{\rm peak}_{\rm dust}$  &  $M_{\rm tot}$    & $N_{\rm H_2}$   & $n$ & CO clump  \\
    
       & ($^h$\ $^m$\ $^s$) & (\gra\ \arcmin\ \arcsec) & (Jy)  & (Jy/beam) & (pc)  & (K)  &  (K)    &  (10$^3$ \msun) & (10$^{22}$ cm$^{-2}$) & (10$^3$ cm$^3$) & counterpart \\   
\hline
\hline
IR1  & 18 16 51.7  & $-$18 41 35     &  14.7  &  0.57   & 1.2  & 33   & 62  & 2.4   & 2.4     &  5.1  &   1 \\
IR2  & 18 16 51.0  & $-$18 43 08     &  3.6   &  0.21   & 1.0  & 27   & 34  & 0.8   & 1.1     &  3.1  &   2  \\
IR3  & 18 16 44.8  & $-$18 40 19     &  7.2   &  0.23   & 1.2  & 25   & 29  & 1.7   & 1.4     &  3.4  &   4  \\
\hline
\end{tabular} 
\label{tabla-IRsources}
\end{table*}

  For the case of clumps 1 and 2, we also obtained the mass and volume density using the HCO$^+$ emission line. Assuming local thermodinamic equilibrium (LTE) the HCO$^+$ column density can be derived from
\begin{equation}
 N({\rm HCO^+}) = 5.85\times 10^{10} e^{(25.7/T_{\rm exc})} \frac{ T_{\rm exc}\ + 0.71}{1 - e^{-(17.12\ / T_{\rm exc}})}\  \int{\tau\ d{\rm v}}
\label{eq:hco}
\end{equation}
 where $\tau$ is the optical depth of the HCO$^+$ line. As excitation temperatures, {\bf $T_{\rm exc}$},  we used a conservative range 20 - 50 K, which is typically used for both low-mass and massive star forming regions. Assuming that the HCO$^+$(4-3) line is optically thin, we can use the approximation
\begin{equation}
\int{\tau\ d{\rm v}\ =\ \frac{1}{J(T_{\rm exc})\ -\ J(T_{\rm bg})}}\ \times\ \int{T_{\rm mb}({\rm HCO^+})\  d{\rm v}} 
\end{equation} 
 where
\begin{equation}
J(T) = \frac{h \nu / k}{e^{(h\nu / kT)}-1}. 
\end{equation}
 being $T_{\rm bg}$ the background temperature ($\sim$ 2.7 K). The LTE mass ($M_{\rm LTE}$, Col. 8 in Table \ref{tabla-propiedades})  can be then obtained assuming a mean fractional abundance X(HCO$^+$)=5$\times$10$^{-11}$ \citep{cor10,cor11} to obtain $N(\rm H_2)$ and using Eq.~\ref{eq:masa} without the correcting factor 0.7.

 Considering  only gravitational and internal pressure, neglecting support of magnetic fields or internal heating sources,  the virialized molecular mass ($M_{\rm vir}$; Col. 9 in Table~\ref{tabla-propiedades})   can  be estimated using the HCO$^+$ line (assumed to be optically thin) from
\begin{equation}\label{eq:virial}
\quad M_{\rm vir}\ =\ {\rm k}\ \  R^{\rm HCO^+}    \ (\Delta{\rm V}^{\rm HCO+})^2  
\end{equation}
\citep{ml88}, where  the constant k depends on the geometry of the density structure, being 126 or 190  according to the density-radius distribution $n \propto r^{-2}$ or $n \propto r^{-1}$, respectively. Virial masses have been derived only for clumps 1 and 2 since no HCO$^+$ emission were detected for clumps 3 and 4.

\section{Submillimeter emission} 

 When an \hii\ region expands, molecular gas and dust accumulate behind  the ionization front, forming shells of dense molecular gas  surrounding the ionization front. Eventually, these shells become massive and could contain cold dust that radiates in the \hbox{(sub-)millimeter} range. The continuum emission at 870 $\mu$m is probably one of the most reliable tracers of the dense molecular gas. It is usually dominated by the thermal emission from dust contained in dense material (e.g. dense star-forming clumps and  filaments) and provides a powerful tool for probing some basic physical properties of dense clumps (column density of molecular hydrogen, mass, etc.) that are needed to unveil the physical conditions in regions where stars can form. 

In  Fig.~\ref{mosaico} (panels {\it b} to {\it i})  we show the 870 $\mu$m emission in green colorscale. Three  bright structures can be discerned centered approximately at  \radec\ =\ 18$^h$16$^m$51.7$^s$, $-$18\gra41\arcmin35\arcsec, \radec\ =\ 18$^h$16$^m$51.0$^s$, $-$18\gra43\arcmin08\arcsec, and \radec\ =\ 18$^h$16$^m$44.8$^s$, $-$18\gra40\arcmin19\arcsec. Their position, size and shape suggest that these sources are the IR counterparts of CO clumps 1, 2, and 4. For the sake of the analysis, we will refer to these  submillimeter sources as IR1, IR2, and IR3 (see panel {\it b} of Fig.~\ref{mosaico}). 

In Table~\ref{tabla-IRsources} we present some physical and geometrical properties derived for the sources. The total (H$_2$ + dust)   mass   of the clumps ($M_{\rm tot}$, Col. 9 in Table \ref{tabla-IRsources}) was calculated from their integrated 870 $\mu$m flux density emission inside the 3 rms level ($S_{870}$, Col. 4 in Table \ref{tabla-IRsources}), assuming that the emission is optically thin, and using the equation 
\begin{equation}
\qquad M_{\rm tot} = R \ \frac{S_{870}\ d^2}{\kappa_{870}\ B_{870}(\bar{T}_{\rm dust})}
\label{masa}
\end{equation}
\citep{hil83},   where $R$ is the gas-to-dust ratio (assumed to be $\sim$ 186; \citealt{jen04,dra07}), $d$ is the distance, $\kappa_{870}$ is the dust opacity per unit mass at 870 $\mu$m  
assumed to be 1.0 cm$^2$ g$^{-1}$  (estimated for dust grains with thin ice mantles in cold clumps; \citealt{osse94}), and $B_{870}(\bar{T}_{\rm dust})$ is the Planck function for a temperature $\bar{T}_{\rm dust}$ (Col. 7 in Table~\ref{tabla-IRsources})  which was estimated averaging the dust temperature map (see below) inside the adopted contour level ($\sim$ 3 rms) over the area $A_{\rm IR}$ corresponding to each source.

 The  average volume density of each clump ($n$, Col. 11 in Table \ref{tabla-IRsources}) was derived, assuming a spherical geometry, as
\begin{equation}
\qquad  n\ =\  \frac{M_{(\rm tot)}}{4/3\ \pi\   R_{\rm IR}^3\ \mu\ m_{\rm H}}  
\end{equation}
where   $m_{\rm H}$ is the mass of the hydrogen atom and $R_{\rm IR}$ (Col. 6 in Table \ref{tabla-IRsources})  is the effective radius of the structure, calculated as $R_{\rm IR} = \sqrt{A_{\rm IR}/\pi}$. The beam-averaged column density of the clumps ($N_{\rm H_2}$, Col.~10 in Table  \ref{tabla-IRsources}) was calculated using
\begin{equation}
\qquad  N_{\rm H_2}\   =\ R\ \ \frac{I_{870}}{\Omega_{\rm beam}\  \kappa_{870}\ \mu\ m_{\rm H}\  B_{870}(\bar{T}_{\rm dust})}
\label{cmd}
\end{equation} 
\citep{hil83}, where $I_{870}$ is the surface brightness emission at 870 $\mu$m (Col.~5 in Table \ref{tabla-IRsources})    and $\Omega_{\rm beam}$ is the beam solid angle \hbox{($\pi$ $\theta_{\rm HPBW}^2$ / 4\ ln(2))}.

 The average and peak temperatures of the submillimeter sources, $\bar{T}_{\rm dust}$ and $T^{\rm peak}_{\rm dust}$, were  estimated using the inverse function of the ratio map of {\it Herschel} 250 and 350\,$\mu$m images, i.e., $T_{\rm dust}=f^{-1}_{(T)}$ (not shown here).  Assuming a dust emissivity following a power law \hbox{$\kappa_{\nu}$ $\propto$ $\nu^{\beta}$},  being ${\beta}$ the spectral index of the thermal dust emission,   in the optically thin thermal dust emission regime $f_{(T)}$ has the parametric form:
\begin{equation}
\qquad f_{(T)} = \frac{S_{250}}{S_{350}} = \frac{B(250,T_{\rm dust})}{B(350,T_{\rm dust})} \left( \frac{250}{350} \right) ^{\beta}
\end{equation}
 where $B(250,T_{\rm dust})$ and $B(350,T_{\rm dust})$   are the blackbody Planck function for a temperature $T_{\rm dust}$ at  the frequencies 250 $\mu$m and 350 $\mu$m, respectively. The temperatures were calculated assuming a typical  value  $\beta$ = 1. The uncertainty in derived dust temperatures using this method was estimated to be about \hbox{$\sim$ 10-15 $\%$}.

\begin{table*}
\caption{Properties derived for the radio continuum sources}
\centering
\begin{tabular}{cccccccc}
\hline
source & RA & Dec.  & $S_{1.4}$   & $n_{\rm e}$    & $EM$ & $M_{\rm ion}$ & $N_{\rm   Lyc}$  \\
    
       & ($^h$\ $^m$\ $^s$) & (\gra\ \arcmin\ \arcsec) & (Jy)  & (cm$^{-3}$)  & (cm$^{-6}$  pc)  & (\msun) & (10$^{47}$s$^{-1}$) \\   
\hline
\hline
A        &  18 16 51.4  &  $-$18 41 29    &  0.09  & 55  & 5000  & 4.6  & 1.0 \\
B        &  18 16 53.7  &  $-$18 37 53    &  0.03  & 30  & 1200   & 3.6  & 0.3  \\
plateau  &   --       &      --         &  0.47  & --  & --     & --  & 5.4\\
\hline
\end{tabular} 
\label{radio}
\end{table*}

The significant higher temperature, derived particulary towards the region corresponding to clump 1,  suggests the existence of powerful internal heating source/s, very likely young stellar objects formed in the region and/or the stellar members of the cluster [BDS2003]\,6 (see Sect. 6.2). 

\section{Radio continuum emission} 

\begin{figure}
   \centering
   \includegraphics[width=250pt]{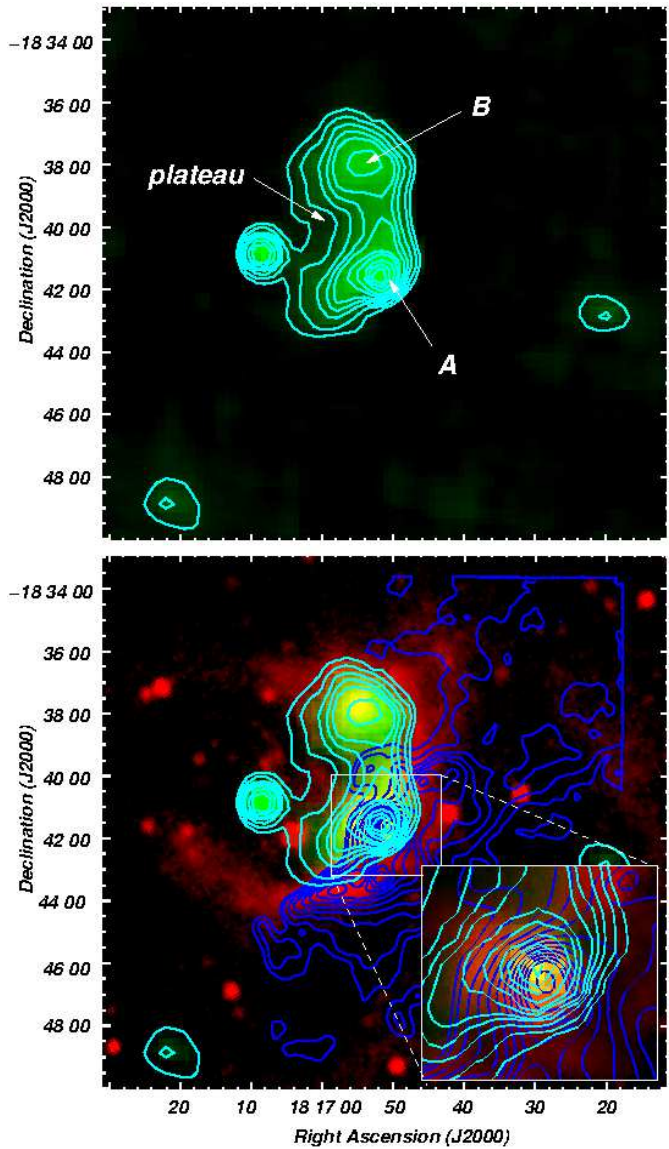} 
   \caption{{\it Upper panel:} Radio continuum emission distribution at 1.4 GHz. Contour levels are 0.0022 ($\sim$ 5 rms), 0.005, 0.01, 0.015, 0.02, 0.03, 0.04, 0.06, 0.08, and 0.1 \mjyb. Identified radio continuum sources are indicated (see text). {\it Lower panel:}  Overlay of the radiocontinuum emission at 1.4 GHz (green color) on the MSX-A emission (red color) and CO emission in the velocity interval from  30 to 45 \kms (blue contours). A zoom of the central region is shown in the lower right corner.}
   \label{nvss}%
    \end{figure}

The radio continuum emission distribution  at 1.4 GHz obtained from NVSS  is shown in the upper panel of Fig.~\ref{nvss}. Two  bright sources can be discerned proyected onto the border of the IR  nebula, which were labeled  as A and B.  The emission peak of  source A is   coincident with the center of the  radio source G012.43-01.1112\  $3.7''\times3.1''$ in size detected at 5 GHz by \citet{urq09}, and  is  almost projected onto the emission peak of the CO clump 1 (see zoomed region in lower panel of Fig.~\ref{nvss}).   The total flux density derived for source A (see below) is in agreement with the value obtained by \citet{bea10} at the same frequency. From the flux density obtained by \citet{urq09}   (S$_{\rm 5 GHz}$ = 0.105 Jy) and the flux density obtained in this work (S$_{\rm 1.4 GHz}$ = 0.09 Jy), we derived  a spectral index $\alpha$ = 0.12 $\pm$ 0.2   (S$_{\nu}$ $\propto$ $\nu^{\alpha}$); using the value derived by \citet{bea10} (S$_{\rm 1.4 GHz}$ = 0.125 Jy) we obtain $\alpha$ = --0.14 $\pm$ 0.2. Both values are  compatible, within the errors, with the thermal free-free regime of an H{\sc ii} region. Then, source A is very likely associated with the  ionization of the molecular gas in clump 1 by the action of  stars possibly belonging  to  the  stellar cluster [BDS2003]6 (see Sect. 6.2).   Source B has a bright IR counterpart at 8 $\mu$m (see Fig.~\ref{msx}) and       is probably associated to IRAS 18139-1839.  A third radio source at 1.4 GHz  is seen projected onto the bubble at \radec\ =\ 18$^h$17$^m$08.4$^s$, $-$18\gra40\arcmin43\arcsec. This source  is  very bright in the VLA MAGPIS image\footnote{https://third.ucllnl.org/gps/index.html} at 0.325 GHz (not shown here).  From its flux densities obtained at 0.325 GHz and 1.4 GHz, we derived  a non-thermal spectral index $\alpha$ = --1.9 $\pm$ 1.7, which indicates that this object is an  extragalactic source.

A noteworthy feature  is an extended horseshoe-shaped structure that connects sources A and B and can be detected almost towards the center of the bubble.  The structure (hereafter the ``plateau'')  shows a  sharp border toward the west of the nebula (see the lower panel of Fig.~\ref{nvss}), while it decreases smoothly  toward the east. This  indicates  the existence of an electron density gradient and suggests that the structure is the radio continuum counterpart of the ionization front and the ionized gas inside the nebula. The morphology of the plateau confirms the proposed scenario of an \hii\ region bounded by ionization to the west (see Sects. 1 and 3.1).  It is worth to point out that neither \citet{urq09} nor \citet{bea10} reported this structure since their MAGPIS VLA images, obtained with more extended VLA configurations, probably were not able to detect extended emission.

   In Table \ref{radio} we present some physical properties estimated for the ionized gas in A, B, and the plateau, assuming optically thin emission at 1.4 GHz.  For the cases of A and B, to avoid the emission contributions  from the plateau,  we use contour line limits of 0.027 \mjyb\ \hbox{($\sim$ 60 rms)} and \hbox{0.021 \mjyb\ ($\sim$ 46 rms)}, respectively. Background-substracted flux densities derived for A and B, and total flux density for the plateau (after substraction of A and B fluxes) are indicated in Col. 4 in Table~\ref{radio}. 

Assuming spherical geometry for A and B, and using the flux densities at 1.4 GHz, the electron density ($n_{\rm e}$, Col 5 in Table  \ref{radio}) was derived using
\begin{equation}
\quad  n_{\rm e} = 3.113\times10^2\ \ S^{0.5}_{1.4}\ \ T^{0.25}_{\rm e}\ \ d^{-0.5}\ \ b(\nu,T_{\rm e})^{-0.5}\ \ \theta^{-1.5}_R\ \  
\end{equation}
where $T_{\rm e}$ is the electron temperature (assumed to be 7 $\times$ 10$^3$ K; \citealt{qui06}), $S_{1.4}$ is the flux density at 1.4 GHz, $\theta_{\rm R}$ is the angular radius of the source, and $b(\nu,T_{\rm e})$ = 1 $+$ 0.3195$\times$log($T_{\rm e}$/10$^4$ K) - 0.213$\times$log($\nu$/1 GHz) \citep{pw78}. The emission measure ($EM$, Col.~6 in Table  \ref{radio}) was obtained using
\begin{equation}
\qquad  EM  = 5.638\times10^4\ \ S_{1.4}\ \ T_{\rm e}\ \ b(\nu,T_{\rm e})\ \ \theta^{-2}_R   
\end{equation}
The ionized gas mass ($M_{\rm ion}$, Col.~7 in Table  \ref{radio})   was derived from
\begin{equation}
\quad  M_{\rm ion}  = 0.7934 \ S^{0.5}_{1.4}\ \ T^{0.25}_{\rm e}\ \ d^{2.5}\ \ b(\nu,T_{\rm e})^{-0.5}\ \ \theta^{1.5}_R\ \ (1+Y)^{-1}  
\end{equation}
where $Y$ is the abundance ratio by number of He$^+$ to H$^+$. The number of ionizing Lyman continuum photons ($N_{\rm   Lyc}$; Col. 8 in Table \ref{radio}) needed to sustain the ionization in the sources can be  also calculated from the 1.4 GHz emission  using
\begin{eqnarray}
\quad    N_{\rm Lyc}\ =\ 7.58\ \times\ 10^{48}\      T^{-0.5}_{\rm e}\  S_{1.4}\ d^2 
\label{nlym}
\end{eqnarray}
\citep{kurtz94}. 
 The physical properties derived for the ionized gas in Table~\ref{radio}   should be considered as lower limits due to the missing flux of VLA. 

  Keeping in mind that values of $N_{\rm   Lyc}$ shown  in Table  \ref{radio}  are lower limits  since about 25 - 50 $\%$ of the UV photons are absorbed by interstellar dust in the H{\sc ii} region \citep{i01}, and taking  into account particulary  the value of $N_{\rm   Lyc}$ derived for the plateau,   we estimate that the total amount of ionizing photons needed to sustain the current level of ionization in Sh2-39 is about 1.1 $\times$ 10$^{48}$ s$^{-1}$. Adopting fluxes  extracted from \citet{ster03}, we  estimate the spectral type of the ionizing star of Sh2-39 (not identified) to be about B0V. Alternatively, a handful of later B-type stars could be also responsible for powering the \hii region.

\section{Star formation} 

In this section we will search for evidence of star formation in the region of Sh2-39. We will use IR catalogs to identify candidate YSOs that are possibly associated with the nebula. We will also pay special attention to the stellar cluster [BDS2003]\,6 and its brightest member, projected at the border of the N5,  which might have  been formed by the action of the nebula on the surrounding gas.

\subsection{Identification of YSOs}

\begin{figure*}
   \centering
   \includegraphics[width=460pt]{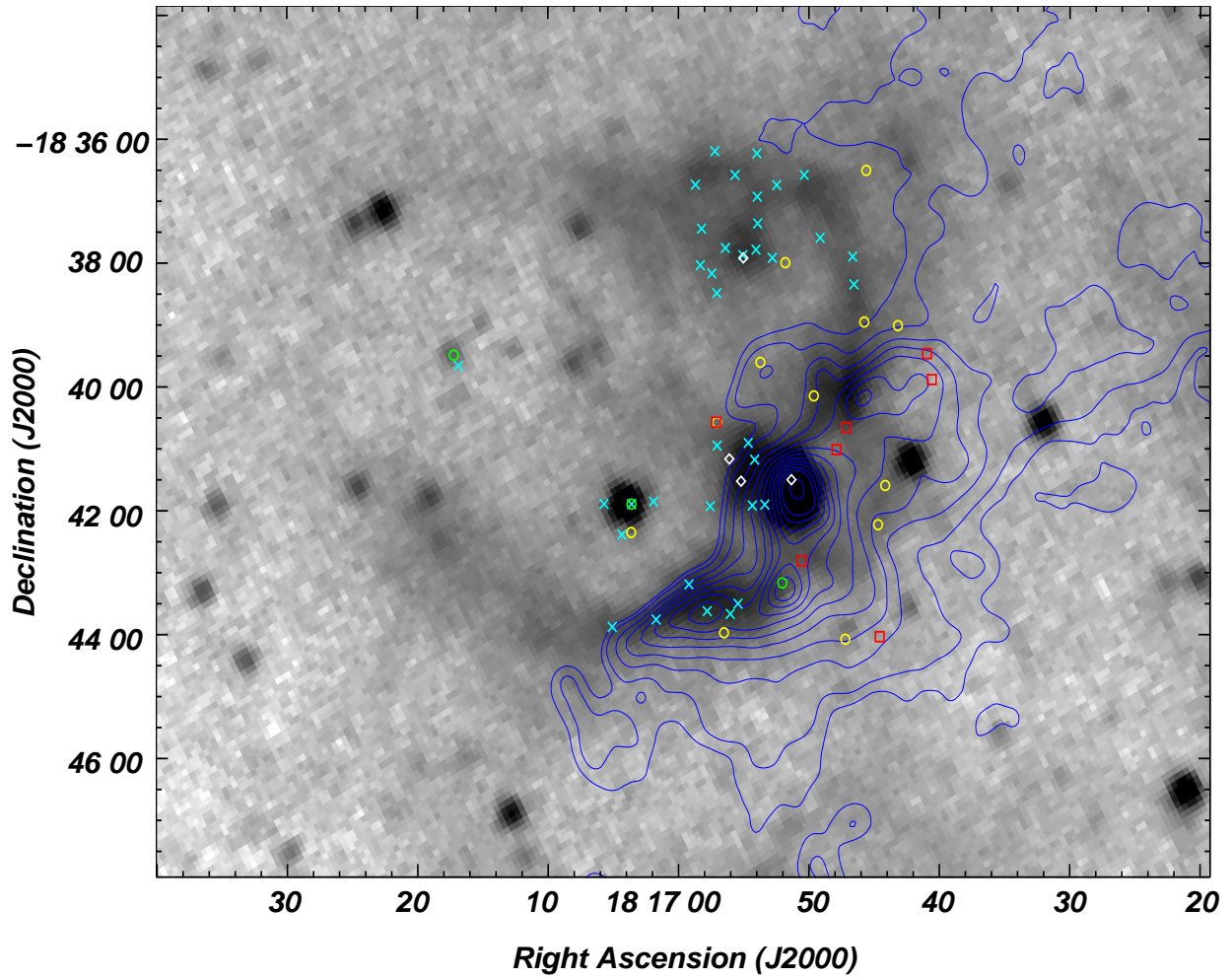} 
   \caption{Candidate YSOs and main sequence late type stars projected onto the MSX image at 8.3 $\mu$m (grayscale) and the CO contours of Fig. 3a. The different symbols have the following meaning: white diamonds: MSX sources; red squares: Spitzer sources; green circles: WISE Class I sources; yellow circles: WISE Class II sources; light blue crosses: Marton et al.'s sourses.}
   \label{ysos-fig}%
    \end{figure*}

 To search for candidate YSOs in N5, we analyzed  the  MSX, WISE, and Spitzer point source catalogs using the color-photometric criteria by \citet{lumsd02}, \citet{kandl14}, and \citet{allen04}, respectively. We searched for protostellar candidates  within a region of 8\arcmin\ in radius centered at the position  \radec\ =\ 18$^h$16$^m$53.1$^s$, $-$18\gra41\arcmin40\arcsec.  We took into consideration only sources projected onto the molecular emission, the IR nebula, and its interior. The coordinates of the identified candidate YSOs, names, and correlations with other catalogs are summarized in Table \ref{ysos}, while their locations are shown in Fig. \ref{ysos-fig}.  In Table \ref{ysos}   we also included the IRAS sources IRAS 18139-1842 and IRAS 18139-1839, which were classified as YSOs by \citet{chan96} and  \citet{codella95}. The source  IRAS 18139-1842 was also identified  as a candidate to be a site of OB star \citep{w&c89}.   
 
 The search for candidates in the MSX catalog  allowed the identification of four sources classified as  massive YSOs (MYSOs) or compact \hii\ region (C\hii). The search in the Spitzer database  allowed to identify 4 Class I sources (protostars surrounded by dusty infalling envelopes) and 3 Class I/II sources (Class II sources: with  emission dominated by accretion disks). Only sources detected in the four Spitzer-IRAC bands were took into consideration. It should be kept in mind  that only part of the nebula  was surveyed with Spitzer. From the WISE database we identified  3 Class I and 13 Class II objects.  For the sake of completeness, in Fig.\ref{ysos-fig} we include 35  ClassI/II candidates extracted from \citet{mart16}, which were identified by the authors using the AllWISE catalog and the support vector machine (SVM) method. 

\begin{figure}
   \centering 
   \includegraphics[width=250pt]{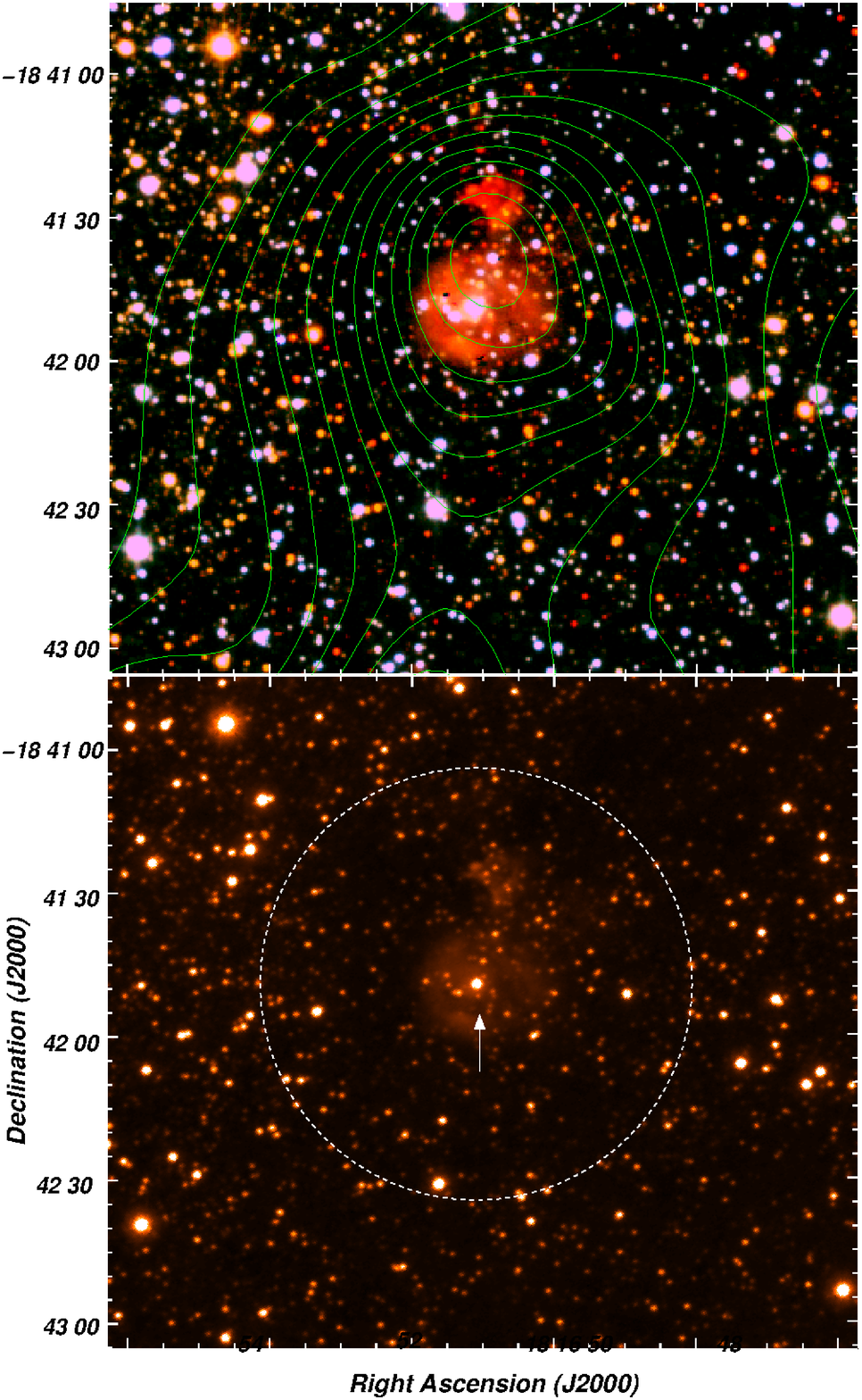} 
   \caption{ {\it Upper panel:} Color (J,H,K1) image of $\sim$ 2\farcm3 $\times$ 2\farcm3  around the center of [BDS2003]\,6 retrieved from the UKIDSSDR10PLUS. Green contours indicate the CO emission of clump 1  {\it Lower panel:} Ks band image of [BDS2003]\,6 retrieved from the VVVX survey.  The radius derived for the cluster (see text)  in shown with the white dotted circle. The location of Obj1 is indicated by the white arrow.  } 
   \label{bds6_rgb}%
    \end{figure}

The spatial distribution of the candidates YSOs  shows that MSX sources 3, 4, and 6 (MYSO candidates), Spitzer source 11 (Class I/II), and a number of Marton et al.'s WISE sources coincide with IRAS18139-1842 (IRAS 1 in Table 5) and the CO clump 1. Radio recombination line observations (H87$\alpha$ and H88$\alpha$)  show emission  towards this source  at a velocity of 41 \kms\ \citep{loc89} which indicates the existence of high density ionized gas at the same radial velocity than the CO,  and is  compatible with the identification of the radio source     G012.4317-01.1112 by \citet{urq09}. The stellar cluster [BDS2003]\,6 is also projected in the region (see Fig.~\ref{bds6_rgb}).    Therefore, the observational evidence suggests that star formation is very active in  the molecular clump 1, which might explain the high peak dust temperature ($\sim$ 60 K) derived from the IR continuum emission in  this region (See Sect.~5). 

Only Spitzer source 12 \hbox{(Class I)} and WISE source 14 \hbox{(Class I)} appear projected onto the CO emission of clump 2. WISE source 26 and a group of Marton et al.'s WISE sources are projected onto CO clump 3, while Spitzer sources 7,  8, and 10, and  WISE sources 17, 21, and 24 seem to be linked to CO clump 4.  Additionally, Spitzer source 9, and WISE  sources 18, 19,  22, and 24 coincide with weak CO emission in the borders of the molecular  region. 


MSX source 5 (MYSO) and a number of  Marton et al.'s WISE sources coincide with IRAS 2 (IRAS18139-1839), as well as the radio continuum source B, indicating that star formation is also active in this region. However, the lack of discernible CO emission within the velocity range linked to the bubble casts doubts on the association of this star forming spot with the \hii\ region. 

Finally, a group of WISE and Marton et al.'s sources is present around  \radec\ =\ 18$^h$17$^m$04.1$^s$, $-$18\gra41\arcmin57\arcsec,  almost projected onto  a bright MSX source. Its relation to the IR bubble is not clear with the present data.

Hence, the presence of many candidate YSOs projected onto the molecular cloud that encircles the western  and southwestern  rims of N\,5 confirms that active star formation has developed in the borders of the bubble,  particulary in the region of clump 1 which is a very active star forming spot.

\begin{table*}
\caption{Identified candidate YSOs projected onto   N\,5.  }
\centering
\begin{tabular}{ccccc}
\hline
\hline
{\bf IRAS sources}     &                                             &                         &                 &         \\
    $\#$               &  RA,Dec                                     &   Catalog name          &   Class         &    \\
                       &   ($^h$\ $^m$\ $^s$, \gra\ \arcmin\ \arcsec) &                         &                 &         \\
\hline
     1                 &  18 16 53.1,  --18 41 40                    &  IRAS18139-1842             &    YSO          &         \\
     2                 &  18 16 53.2,  --18 37 57                    &  IRAS18139-1839             &    YSO          &         \\
\hline
\hline  
{\bf MSX sources}      &                                             &                         &                 &  Matching with      \\
    $\#$               &                                             &                         &                 &  other sources       \\
\hline
     3                 &  18 16 56.1,  --18 41 09           &  G012.4452-01.1255       & MYSO/C\hii     &          \\
     4                 &  18 16 55.2,  --18 41 32                    &  G012.4382-01.1252       & MYSO           &          \\
     5                 &  18 16 55.0,  --18 37 55                    &  G012.4909-01.0962       & MYSO           &  {\it c}          \\
     6                 &  18 16 51.3,  --18 41 30                    &  G012.4314-01.1117       & MYSO           &          \\   
\hline
\hline
{\bf Spitzer sources}  &                                             &                         &                 &         \\
    $\#$               &                                             &                         &                 &         \\
\hline
    7                  & 18 16 40.583,  --18 39 53.02                & G012.4350-01.0616       &   I             &         \\
    8                  & 18 16 40.954,  --18 39 27.94                & G012.4418-01.0596       &   I             &       {\it a} \\
    9                  & 18 16 44.566,  --18 44 02.22                & G012.3816-01.1083       &   I             &       {\it a}  \\
   10                  & 18 16 47.132,  --18 40 39.39                & G012.4360-01.0905       &   I/II          &           \\
   11                  & 18 16 47.887,  --18 41 01.02                & G012.4321-01.0959       &   I/II          &           \\
   12                  & 18 16 50.554,  --18 42 48.80                & G012.4107-01.1194       &   I             &           \\
   13                  & 18 16 57.106,  --18 40 34.36                & G012.4559-01.1244       &   I/II          & 27         \\
\hline
\hline
{\bf WISE sources}     &                                             &                         &                 &         \\
    $\#$               &                                             &                         &                 &         \\
\hline
   14                  & 18 16 52.045, --18 43 10.16                 & J181652.04-184310.1     &  I              &         {\it b} \\
   15                  & 18 17 03.605, --18 41 53.73                 & J181703.60-184153.7     &  I              &         {\it c} \\
   16                  & 18 17 17.233, --18 39 29.28                 & J181717.23-183929.2     &  I              &         {\it c} \\
   17                  & 18 16 43.158, --18 39 01.24                 & J181643.15-183901.2     &  II             &         {\it b} \\
   18                  & 18 16 44.147, --18 41 35.80                 & J181644.14-184135.8     &  II             &         {\it b} \\
   19                  & 18 16 44.702, --18 42 13.71                 & J181644.70-184213.7     &  II             &         {\it b} \\
   20                  & 18 16 45.609, --18 36 30.91                 & J181645.60-183630.9     &  II             &         {\it b} \\
   21                  & 18 16 45.762, --18 38 57.27                 & J181645.76-183857.2     &  II             &         {\it b} \\
   22                  & 18 16 47.073, --18 44 03.27                 & J181647.07-184403.2     &  II             &          \\
   23                  & 18 16 49.622, --18 40 08.44                 & J181649.62-184008.4     &  II              &         {\it b}  \\ 
   24                  & 18 16 51.811, --18 37 57.27                 & J181651.81-183757.2     &  II             &         {\it b}  \\
   25                  & 18 16 53.809, --18 39 36.41                 & J181653.80-183937.4     &  II             &         {\it b}  \\
   26                  & 18 16 56.510, --18 43 58.57                 & J181656.50-184358.4     &  II             &           \\
   27                  & 18 16 57.048, --18 40 34.64                 & J181657.04-184034.6     &  II             &          13\\
   28                  & 18 17 03.605, --18 41 53.73                 & J181703.60-184153.7     &  II             &           \\
\hline
\hline

\end{tabular} 
\tablefoot{{\bf :}\\{\it a}: coincidence with WISE source not cataloguedd as YSO \\ {\it b}: coincidence with Spitzer source not catalogued as YSO \\ {\it c}: Identified by \citet{mart16} as YSO Class II}
\label{exc-candidates}
\label{ysos}
\end{table*}

\subsection{The star cluster [BDS2003]\,6 and 2MASSJ18165113-1841488}
 
As mentioned in Sect. 2,  \citet{bica03} identified a small star cluster candidate, with nomenclature [BDS2003]\,6 in their catalog, in the region of Sh2-39.  Later on, \citet{mor13} estimated  a distance of 4.15 kpc to the cluster. As previously mentioned, the cluster is projected onto the region of CO clump 1, a very active center of star formation.
 \begin{figure}
   \centering
   \includegraphics[width=190pt]{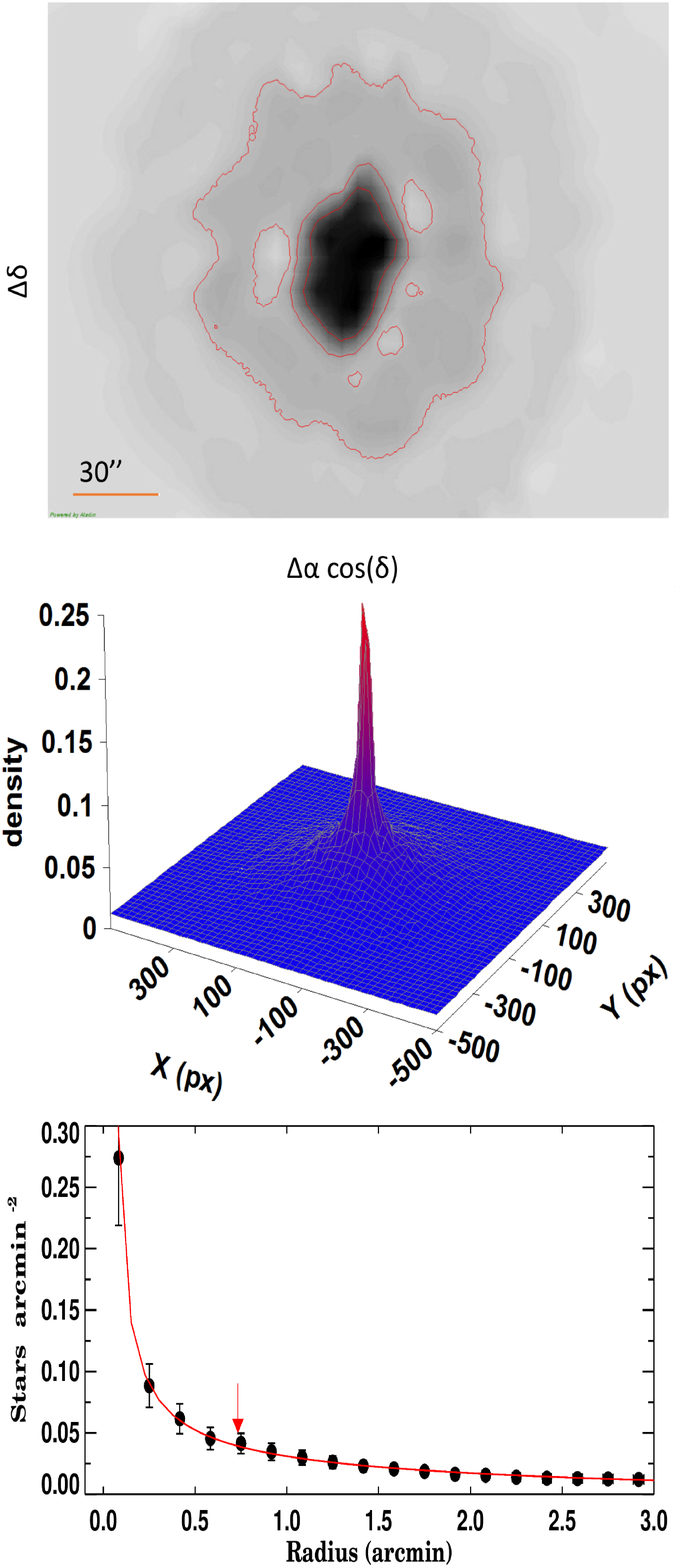} 
   \caption{{\it Upper panel:} 2D stellar surface-density map obtained in a region 3$'$ around the  center of [BDS2003]\,6.  The red lines corresponds to the contours representing 3, 5 and 10 sigma of the background surface density. {\it Middle panel:} 3D stellar surface-density map. {\it Lower panel:} Radial density profile of the cluster. The solid line stands for the best fit and the arrow marks the radius derived for the cluster.     }

   \label{profiles} 
    \end{figure}  
  \begin{figure}
   \centering
   \includegraphics[width=260pt]{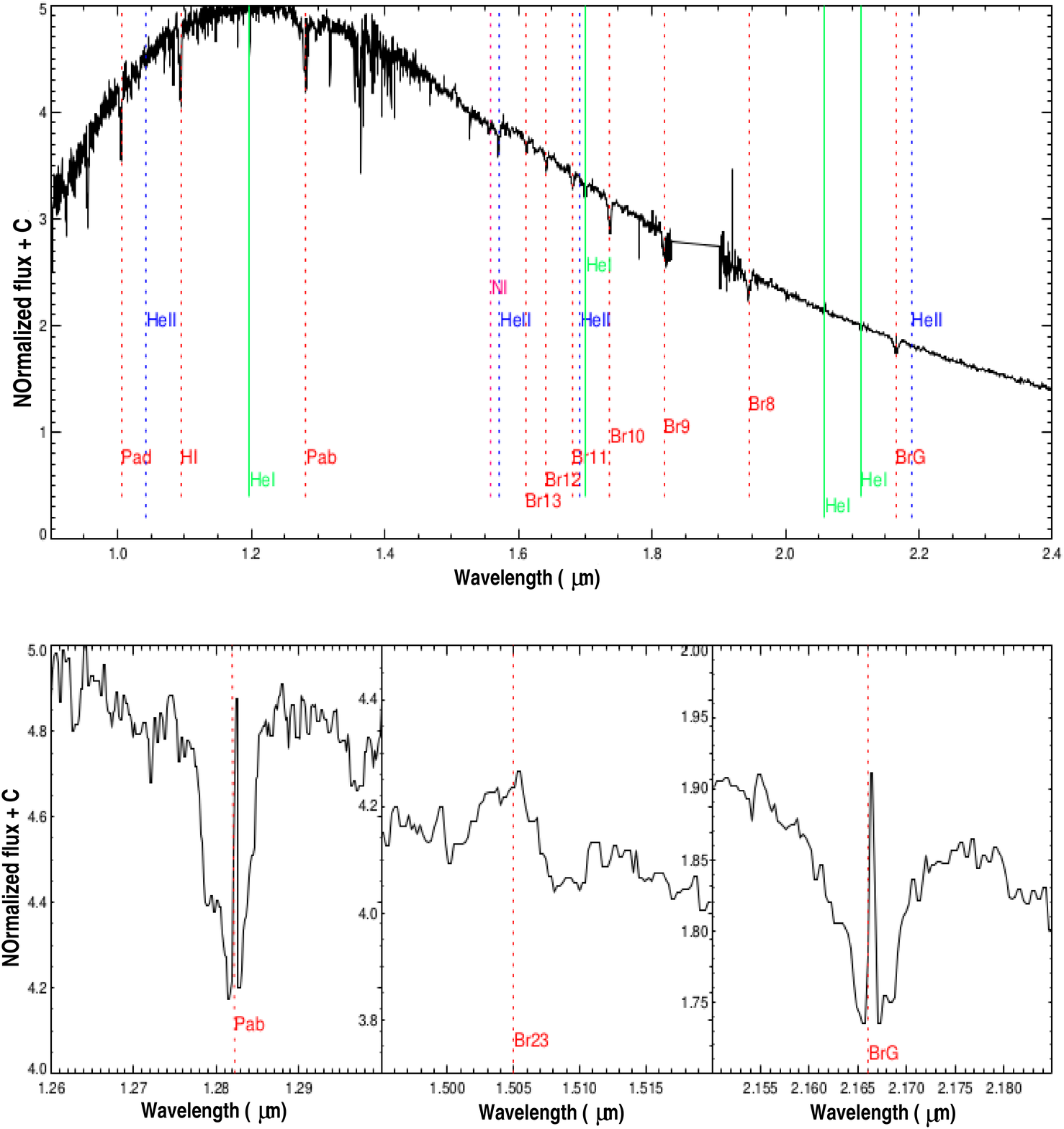} 
   \caption{The ARCoIRIS spectrum of  Obj1 (2MASSJ18165113-1841488). The observed spectral lines are marked. Lines with emission components are zoomed.}
   \label{obj1_sp} 
    \end{figure}  

 The color (J,H,K1) image of $\sim$ 2\farcm3 $\times$ 2\farcm3   around the center of [BDS2003]\,6 retrieved from UKIDSS is shown in the upper panel of Fig.~\ref{bds6_rgb}.  By an aperture photometry performed on these bands, only a handful of stars of a poorly populated main-sequence (MS) and pre-main-sequence (PMS) were identified in  the color-color and color-magnitude diagrams  as probably members of the cluster. They have magnitudes in the range 10-16 mag in Ks and  colors in the  range 1.3 $<$ (J$-$Ks) $<$ 4.5. No further analysis was performed on these data due to small number of stars in the UKIDSS photometric catalog. 

 In the lower panel of Fig.~\ref{bds6_rgb} we show the Ks band image of [BDS2003]\,6 obtained  from the VVVX survey. We have performed a  PSF photometry transforming magnitudes to the 2MASS system using common stars (see  \citealt{bori11,bori14} for details of the procedure).  The photometry was used to construct the stellar surface-density map of the cluster and to determine its visual radius.  We a performed direct star counting with a 10 arcsec space radius, assuming spherical symmetry. That number is then divided by the area of the rings to determine the stellar density.  The obtained spatial distribution map of the stellar surface density is shown in the upper and middle panels of Fig.~\ref{profiles}. The density peak is about 10 times larger than the surface density of the comparison field (estimated to be 0.025 $\pm$ 0.009 stars/arcmin$^2$), which confirms the clustering nature of  [BDS2003]\,6. To estimate the size of the cluster, we used the projected star surface density as a function of the radius (lower panel of Fig.~\ref{profiles}) and  the theoretical profile  of \citet{elson87}. We obtained a visual radius of  44$''$$\pm$8$''$ (0.87 pc at a distance of 4.1 kpc), almost twice the radius estimated by \citet{mor13}, and approximately the same size of clump 1.       A quantitative estimate of the cluster compactness can be obtained from the density-contrast parameter, $\delta$,  proposed by \citet{bobi09}, which connects the surface density of the cluster, $\sigma_{\rm cl}$, with the residual background density, $\sigma_{\rm bg}$, by  $\delta$ =  $\sigma_{\rm cl}$/$\sigma_{\rm bg}$ + 1. For the case of [BDS2003]\,6 we estimated $\delta$ = 11.8, which represents a relatively compact cluster.

To estimate the reddening and the distance to the cluster we use the spectroscopic parallax of its brightest source, 2MASSJ18165113-1841488 (hereafter Obj1), observed with ARCoIRIS. The position of this source is indicated in the lower panel of Fig.\ref{bds6_rgb}. Spectral classification was performed using atlases of infrared spectra that feature spectral types stemming from optical studies (e.g \citealt{han05}). The reduced spectrum is shown in Fig.~\ref{obj1_sp}, where  He\,I, Paschen, and Brackett lines in absorption are clearly visible, but no He\,II is observed. In addition, we take in consideration the position of Obj1 on the color-magnitude diagram (not shown here) to discern about its luminosity class. Gathering all this information, we  derive a spectral type O8-9Ve.  It is worth mentioning that the profiles in the lines  P$\beta$,  Br\,23, and Br$\gamma$ are partly filled  (see lower panel of Fig.~\ref{obj1_sp}), which is believed to  be caused by an enhanced mass-loss episode, like stellar winds or molecular outflows  \citep{bik08}. The presence of a very bright  shell-like nebulosity around Obj1  (see upper and lower panels of Fig.~\ref{bds6_rgb}) provides a strong support to this scenario.  A second nebulosity is seen  $\sim$ 25$''$ north of Obj1 at \radec\ =\ 18$^h$16$^m$50.8$^s$, $-$18\gra41\arcmin27\arcsec,    having the same size and position than the radio source G012.4317-01.1112 reported by \citet{urq09}. This coincidence suggests that this nebulosity could be the IR counterpart of radio source A.  A higher spatial resolution study of these nebulosities might provide important information on the characteristics of Obj1 and the rest of the stellar content of the cluster. 

Individual extinction and distance were estimated  using the spectral classification of  Obj\,1 and the intrinsic colors and luminosities cited by \citet{mar06} for O type stars. The uncertainties are calculated by quadratically adding the uncertainties of the photometry and the spectral classification (e.g., 2 subtypes).  We derived E(J-K) = 1.6 $\pm$ 0.3 and $(M-m)_0$ = 13.1 $\pm$ 0.43, which corresponds to a distance  $d$ = 4.2 $\pm$ 0.8 kpc. This value is  in excellent agreement with the distance estimated by \citet{mor13} and the   kinematical distance adopted for Sh2-39. This confirms undoubtedly   that Obj1 and the stellar cluster [BDS2003]\,6 are physically associated with N5.  The high dust temperatures derived in the region of clump 1 (see Sect. 4) could be explained by the presence of the cluster, still embedded in the the dense molecular gas of the clump.    The formation of the cluster  was very likely triggered by the expansion of the ionization front over its parental molecular cloud. A discussion of this scenario is presented in Sect.~7.3.   


\section{Discussion}

\subsection{Stability  of the clumps}

 Classical Virial equilibrium analysis establishes that for a stable, self gravitating spherical mass, the kinetic energy must equal the half of the potential energy. Then, the  virial mass   estimated  with Eq.~\ref{eq:virial}    is the minimum mass required in order for a clump to be gravitationally bound. This  means that  if $M_{\rm vir}$  is smaller  than the mass estimated with any other technique, the clump is gravitationally unstable and has potential to collapse to for new stars. In the same way, if $M_{\rm vir}$ is larger,  the clump has too much kinetic energy and is unable to collapse. 

  Considering the range of molecular masses detailed in Tables~\ref{tabla-propiedades} and \ref{tabla-IRsources} for clump 1, we estimate the  mass ratio ranges  $M_{\rm int}$/$M_{\rm vir}$ = 0.9 - 1.5 and  $M_{\rm LTE}$/$M_{\rm vir}$ = 0.6 - 2.5, which suggests a more likely  gravitational collapse. For the case of the ratio $M_{\rm tot}$/$M_{\rm vir}$ = 1.1 - 1.8 the result is more conclusive. As seen before, clump 1 seems to be a very active massive star forming clump, which confirms that the clump  is, or else was undergoing gravitational collapse.  For the case of clump 2, where two candidate YSOs were found projected onto, the ratios are $M_{\rm int}$/$M_{\rm vir}$ = 2.4 - 3.6 and  $M_{\rm LTE}$/$M_{\rm vir}$ = 0.6 - 2.5, and  $M_{\rm tot}$/$M_{\rm vir}$ = 0.9 - 1.3, which also suggests a likely gravitational collapse.

 The mass-radius relationship of nearby molecular clouds was investigated in two sequent papers by \citet{kauf10a} and \citet{kauf10b}  who found that clouds that were devoid of any high-mass star formation ussually obeyed the empirical relationship $m(r) \le\ 870 $\msun$ \times\ (R_{\rm eff}/{\rm pc})^{1.33}$. Afterwards, \citet{urq13} studied 577 submillimeter continuum clouds and determined a lower surface density of 0.05 gr/cm$^{-2}$  (M $>$ 750 \msun $\times$  ($R_{\rm eff}$/pc)$^2$) as a reliable empirical lower bound for the clump surface density required for massive star formation. Using $R_{\rm eff}$ derived for clumps 1 (see Table \ref{tabla-propiedades}) we obtain masses of 1.3 $\times$ 10$^3$ \msun and 1.5 $\times$ 10$^3$ \msun, for the Kauffmann et al.'s and Urquhart et al.'s relations, exceeding  both mass limit conditions for the massive star formation.  For the case of clump 2 we obtain masses of 2.0 $\times$ 10$^3$ \msun and 2.7 $\times$ 10$^3$ \msun, exceeding only  Kaufmann et al.'s limit mass. 

Since no HCO$^+$ emission could be observed for clumps 3 and 4, a virial analysis was not possible to perform. Effective radius estimated for these clumps indicate that both clumps have masses in the limit of the Kauffmann et al.'s masses, but are under the Urquhart et al.'s limit. As seen in Sect. 8, star formation activity seems to be very active in clump 3, which may indicate that a gravitational collapse may have also ocurred here.

\subsection{The expansion of the nebula}

In previous sections, we show observational evidence suggesting that the ionized nebula is expanding against the molecular cloud. Although no conclusive evidence of expansion can be obtained from molecular data,  the dynamical time, $t_{\rm dyn}$, and the expansion velocity, $\dot{R}_{\rm HII}$, can be obtained  from the radio continuum emission by using the model of \citet{dw97}. The radius of  an \hii region ($R_{\rm HII}$)  in a uniform medium is given by
\begin{equation}
\quad  \frac{R_{\rm HII}}{R_{\rm S}}\ =\ \left(1\ + \frac{7\ {\rm v}_{\rm s}\ {t}_{\rm dyn}}{4\ R_{\rm S}}\right)^{4/7} 
\label{dw}
\end{equation}
where v$_{\rm s}$ is the sound speed in the ionized gas \hbox{(assumed to be $\sim$ 10 \hbox{\kms})}, and    $R_S$ is the radius of the Str\"omgren sphere  \citep{s39} before expanding, given by 
\begin{equation}
\quad  R_{\rm S} = \left(\frac{ 3 N^*_{Lyc}}{4 \pi\ (n_0)^2 \alpha_{\beta}} \right)^{1/3} 
\end{equation}
where  $n_0$ is the original ambient density, and $\alpha_{\beta}$ is the  hydrogen recombination  coefficient to all levels above the ground level (adopted as 3.5 $\times$ 10$^{-13}$ for $T_e$= 7000; \citealt{leq05}).  To estimate $n_0$ we averaged the total mass of the molecular clumps \hbox{($\sim$ 7.5 $\times$ 10$^3$ \msun)} over a sphere 4.8 pc  in radius ($\sim$ 4$'$ at a distance of 4.1 kpc), which yields to  $n_0$ =  620 cm$^{-3}$. Adopting  $N^*_{Lyc}$ =  1.1 $\times$ 10$^{48}$ s$^{-1}$  (see Sect. 5), we infer \hbox{$t_{\rm dyn}$ $\approx$ 8.4 $\times$ 10$^6$} yr. We keep in mind that when considering only the mass of the molecular clumps, and neglecting extended molecular emission, a considerable  amount of molecular mass have not been taken into account, which yields to a lower limit for   $n_0$. Therefore, $t_{\rm dyn}$ could be even larger than the value derived above.

The expansion velocity of the \hii region  can be estimated by deriving Eq.\ref{dw} in time   
\begin{equation}
\quad  \dot{R}_{\rm HII} = v_s \left( \frac{R_{\rm HII}}{R_s} \right)^{-3/4}
\end{equation}
 yielding to an expansion velocity  of about  $\sim$ 0.4  km s$^{-1}$.  

The obtained dynamical age and expansion  velocity indicates that Sh2-39 is older and slower compared  with  typical Galactic bubble \hii regions (e.g. Gum 31, \citealt{cn08}; Sh2-173, \citealt{ci09}; G8.14+0.23, \citealt{dewa12}; RCW 120, \citealt{and15}).

In Sect. 5 we estimated the spectral type of the ionizing star/s in Sh2-39 to be BOV or later. The total  main sequence lifetime for these spectral types  is estimated to be  $\sim$  10$^7$ yr \citep{han94}, which is in agreement with the dynamical time derived for Sh2-39.

\subsection{Triggered star formation scenario}

To test whether the C$\&$C mechanism is responsible for the formation of [BDS2003]\,6  and the candidate YSOs identified in the periphery of Sh2-39, we compare the dynamical age of the \hii\ region, $t_{\rm dyn}$,   with the fragmentation time of the collected molecular gas, $t_{\rm frag}$,  predicted with the model by  \citet{wi94}.

To estimate $t_{\rm frag}$, and  the size of the \hii region at the fragmentation time ($R_{\rm frag}$), we use
\begin{equation}
\quad t_{\rm frag} [10^6 yr]\ =\ 1.56\ a_{0.2}^{4/11}\ n_3^{-6/11}\ N_{49}^{-1/11}  \ [10^6 \ \ {\rm yr}]
\end{equation}
\begin{equation}
\quad R_{\rm frag} [pc]\ =\ 5.8\  a_{0.2}^{4/11}\ n_3^{-6/11}\ N_{49}^{1/11}   \ \ [{\rm pc}]
\end{equation}
 where $a_{0.2}$ is the isothermal sound speed in the compressed layer, in units of 0.2 \kms\ ($a_s/0.2$ \kms), $n_3$ is the  mean density of the surrounding homogeneous infinite medium into which the \hii\ region expands, in units of 1000 cm$^{-3}$ ($n_0/1000$ cm$^{-3}$), and $N_{49}$  is the number of ionizing photons, in units of 10$^{49}$ ($N_{49} =  N_{Lyc}^*/10^{49}$ s$^{-1}$). As pointed out in   \citet{wi94},  $a_{s}$ = 0.2  \kms\  is  likely a lower limit for the sound of speed in the collected layer, since both turbulence and extra heating from intense sub-Lyman-continuum photons leaking from the \hii\ region could increase this value. Therefore, in the following we will use the typically adopted  range $a_{s}$ =  0.2 - 0.6 \kms\   for the collected layer (e.g \citealt{dewa12}).  We obtained a fragmentation time range \hbox{$t_{\rm frag}$ = 2.5 - 3.6 $\times$ 10$^6$  yr}, and  a frangmentation radius range  \hbox{$R_{\rm frag}$ = 6.2 - 9.3 pc}. A direct  comparison between $t_{\rm dyn}$ and $t_{\rm frag}$ suggests that C$\&$C is a very plausible scenario for the star formation in the border of N5. 
However, a comparison between $R_{\rm HII}$ and $R_{\rm frag}$ suggests that the fragmentation should have occured some later.  A multiple mechanism scenario (C$\&$C-RDI-gravitational instabilities) is also possible. More observational data are necessary to shed some ligth on these issues. 

We keep in mind that some discrepancies are always expected between model and observations, namely: a) the model of \citet{wi94} assumes spherical symmetry around the central star   and  expansion into an homogeneous infinite medium. This is certainly not the case of Sh2-39, which appears to have been formed  at the edge of its  parental cloud (blyster-type \hii\ region). b) if the fragmentation had symmetry around the center of Sh2-39, more  candidate YSOs should be detected in the direction of the ionized gas. c) although the ring morphology in the IR  emission is a common feature  in  many  Galactic \hii\ regions  (e.g. \citealt{wat08,deh09,ca16}) it  has originated some debate on whether these structures are flat (few parsecs) two-dimensional ring-like objects \citep{bea10,arce11} or projected three-dimensional bubbles \citep{deh10,and12,and15,du15}.

\section{Summary and conclusion}
 
As  part of a project aimed to study the  physical characteristics of Galactic IR bubbles and to explore their influence in triggering  massive star formation, we present a multiwavelength analysis of the northern \hii\ region bubble Sh2-39 (N5)  and its environs. We used ASTE CO(3-2) and HCO$^+$(4-3) lines data to explore the characteristics of the molecular cloud associated with the nebula and to study the physical properties of the star forming locations. The characteristics of the dust and ionized gas were explored using archival data, while a stellar source member of the cluster [BDS2003]\,6, projected onto the border of the IR bubble, was studied using new IR data. Our findings are summarized as follows:

\begin{itemize}

\item  Using the CO(3-2) line data, we  identified a molecular cloud  concentrated between $\sim$ 30 \kms\ and 46 \kms\ towards the western and southern borders  of the IR nebula, coincident with the PAH emission detected at 8.28 $\mu$m, indicating that Sh2-39 is a blister-type \hii\ region bounded by ionization to the west and by density to the east.  \\

\item  We identified four  molecular clumps having radii between 1.4 and 1.9 pc, integrated masses between 1.4 and 2.2 \msun, and volume densities between 1.5 and 3.3 \cm3.  The clumps are located over the western border of the IR bubble and were likely formed by the action  of the \hii\ region expanding against its parental molecular cloud. Clumps 1 and  2 have counterparts in the  HCO$^+$(4-3) line and  in the 870 $\mu$m dust continuum emission, while clump 3 only in the 870 $\mu$m emission. This indicates that these clumps have high densities and are good sites to look for massive star formation. A virial analysis performed for clumps 1 and 2 indicates that both clumps are gravitationally unstable and have potential to collapse to for new stars.   \\

\item Using IR point surce catalogs, we identified a number of candidate YSOs projected onto the molecular cloud next to N5 confirming that active star formation has developed in the borders of the bubble. The location of the stellar cluster [BDS2003]\,6, projected onto the molecular clump 1, is also indicative of the star formation activity.  An analysis performed using new available IR data indicates that the cluster is 0.87 pc in radius and confirms its  clustering nature.  A spectroscopic study of the  brightest member of   [BDS2003]\,6  (Obj1) reveals a O8-9V star. The estimated the distance to the cluster, 4.2 $\pm$ 0.8 kpc, is in excellent agreement with previous distance estimations for Sh2-39. This certainly confirms the association of the cluster  with the nebula and gives more support to the massive star formation scenario on the edge of the \hii\ region,  particulary on clump 1.\\


\item The analysis of the  NVSS radio continuum emission distribution at 1.4 GHz allowed us to identify three noteworthy features projected towards the center and the border of the IR bubble. The morphlogy of the more extended feature (plateau) perfectly matches the inner border of the PDR which indicates that this structure is the radio counterpart of the ionization front and the \hii\ region.  We estimate a single B0V, or else a handful of later B-type stars to be reponsible for powering the \hii\ region.  We derived a dynamical age for the nebula of  $\sim$ 8.4 $\times$ 10$^6$ yr, wich is in agreement with the  main sequence lifetime of a B0V star or later. The dynamical age and expansion velocity derived for Sh2-39 indicates that the bubble is older and slower compared  with  typical Galactic bubble \hii\ regions.     The other two radio sources (A and B) were probably formed during the expansion of Sh2-39.\\

\item After comparing the dynamical age of the \hii\ region and the fragmentation time of the molecular gas around the \hii\ region, we  have concluded  that the collect-and-collapse process  may indeed be important in the collected layers of gas at the edge of the bubble, boosting the formation of the candidate YSOs detected in the molecular gas around Sh2-39, as well as the cluster [BDS2003]\,6. However, a comparison between the observed radius of the bubble with the fragmentation radius suggests that the fragmentation should be occurred later.

\end{itemize}

\begin{acknowledgements}
  We acknowledge the anonymous referee for her/his helpful comments and suggestions that improved the presentation of this paper. This project was partially financed by CONICET of Argentina under projects PIP 00356, and  PIP00107,  and  from UNLP, 2012-2014 PPID/G002 and  11/G120.  L.B. acknowledges support by CONICYT project PFB-06.  Support for JB  is provided by the Ministry of Economy, Development, and Tourism's Millennium Science Initiative through grant IC120009, awarded to The Millennium Institute of Astrophysics, MAS.

\end{acknowledgements}

%
%

\bibliographystyle{aa}
\bibliography{bibliografia-sh2-39}
 
\IfFileExists{\jobname.bbl}{}
{\typeout{}
\typeout{****************************************************}
\typeout{****************************************************}
\typeout{** Please run "bibtex \jobname" to optain}
\typeout{** the bibliography and then re-run LaTeX}
\typeout{** twice to fix the references!}
\typeout{****************************************************}
\typeout{****************************************************}
\typeout{}

}


\end{document}